\documentclass[num-refs]{wiley-article}

\usepackage{siunitx}
\usepackage{subfig}
\usepackage{wrapfig} 
\usepackage{textcomp}
\usepackage{comment}
\usepackage{microtype}
\usepackage[linesnumbered, ruled, boxed]{algorithm2e}
\usepackage{graphicx}
\usepackage{epstopdf}
\usepackage{listings}
\usepackage{multirow}
\usepackage{url}
\usepackage{xcolor}
\usepackage{verbatim}
\usepackage{balance}
\usepackage{pifont}
\usepackage{enumitem}
\usepackage{adjustbox}
\usepackage{textcomp,booktabs}
\usepackage[normalem]{ulem}
\usepackage{caption}
\usepackage{color}
\usepackage{bm}
\usepackage{tcolorbox}
\usepackage[T1]{fontenc}
\usepackage{threeparttable}
\usepackage{tikz}
\usepackage{diagbox}
\usepackage{hyperref}
\usepackage{colortbl}

\papertype{Research Article - Empirical}

\title{Revisiting Deep Neural Network Test Coverage from the Test Effectiveness Perspective}

\author[1]{Ming Yan}
\author[1]{Junjie Chen}
\author[1]{Xuejie Cao}
\author[2]{Zhuo Wu}
\author[1]{Yuning Kang}
\author[1,2]{Zan Wang}

\affil[1]{College of Intelligence and Computing, Tianjin University, Tianjin, Tianjin, 300350, China}
\affil[2]{School of new media and communication, Tianjin University, Tianjin, Tianjin, 300072, China}

\corraddress{Junjie Chen, College of Intelligence and Computing, Tianjin University, Tianjin, Tianjin, 300350, China}
\corremail{junjiechen@tju.edu.cn}

\runningauthor{Yan et al.}

\fundinginfo{National Natural Science Foundation of China, Grant/Award Number: 62232001 and 62002256}

\newcommand{\jj}[1]{{\color{orange}\ding{46}[junjie: #1]}}
\newcommand{\ming}[1]{{\color{cyan}\ding{46}[ming: #1]}}

\newcommand{\yes}{\ding{51}}
\newcommand{\no}{\ding{55}}

\definecolor{MyGreen}{cmyk}{0.92,0,0.87,0.09}
\definecolor{MyGreen-gray}{gray}{0.5383}
\definecolor{MyGreen-cmy}{cmy}{1,0.09,0.95999}
\DeclareRobustCommand{\legendsquare}[1]{%
	\textcolor{#1}{\rule{4ex}{2ex}}%
}

\begin{document}

\begin{frontmatter}
\maketitle

\begin{abstract}
Many test coverage metrics have been proposed to measure the Deep Neural Network (DNN) testing effectiveness, including structural coverage and non-structural coverage.
These test coverage metrics are proposed based on the fundamental assumption: they are correlated with test effectiveness.
However, the fundamental assumption is still not validated sufficiently and reasonably, which brings question on the usefulness of DNN test coverage. 
This paper conducted a revisiting study on the existing DNN test coverage from the test effectiveness perspective, to effectively validate the fundamental assumption.
Here, we carefully considered the diversity of subjects, three test effectiveness criteria, and both typical and state-of-the-art test coverage metrics.
Different from all the existing studies that deliver negative conclusions on the usefulness of existing DNN test coverage, we identified some positive conclusions on their usefulness from the test effectiveness perspective.
In particular, we found the complementary relationship between structural and non-structural coverage and identified the practical usage scenarios and promising research directions for these existing test coverage metrics.



\end{abstract}
\end{frontmatter}

\section{Introduction}\label{sec:intro}
\vspace{-3mm}
Deep neural networks (DNNs) have been widely studied in recent years and have achieved great success in many domains, e.g., autonomous driving cars~\cite{chen2015deepdriving}, face recognition~\cite{sun2014deep}, 
machine translation~\cite{cheng2019semi}, 
medical diagnosis~\cite{obermeyer2016predicting}, 
and code analysis~\cite{DBLP:conf/icse/GuZ018,DBLP:conf/sigsoft/GuZZK16,kang2021APIRecX,tian2022learning}.
However, like traditional software systems, DNN also contains bugs~\cite{pei2017deepxplore,ma2018deepgauge,odena2018tensorfuzz}.
DNN bugs could lead to unexpected behaviors, even disasters in safety-critical domains.
For example, an Uber autonomous driving car killed a pedestrian in Tempe, Arizona in 2018~\cite{news1}.
Therefore, it is very critical to ensure the DNN quality.

DNN testing is one of the most effective methods to ensure the DNN quality.
As with traditional software testing~\cite{inozemtseva2014coverage,zhang2015assertions,DBLP:conf/icse/ChekamPTH17,DBLP:conf/saicsit/MorrisonIV12,DBLP:conf/kbse/Hilton0M18,DBLP:conf/issta/GligoricGZSAM13,DBLP:journals/tosem/GayRSWH16,DBLP:journals/tosem/GligoricGZSAM15}, one important aspect in DNN testing is to measure test effectiveness.
Indeed, many DNN test coverage metrics have been proposed to achieve this goal in recent years, e.g., neuron coverage measuring which DNN elements are covered during the testing process~\cite{pei2017deepxplore,ma2018deepgauge} and surprise coverage measuring the difference of DNN behaviors between test inputs and training data~\cite{kim2019guiding}.
All these test coverage metrics are proposed based on the same assumption: \textit{the proposed test coverage metric is correlated with test effectiveness}.

However, this assumption is not sufficiently validated till now, which also brings question on the usefulness of DNN test coverage.
Although some experiments have been conducted when these test coverage metrics were proposed, they ignored the influence of some reasonable factors such as test-set sizes.
For example, in the existing experiments~\cite{ma2018deepgauge,kim2019guiding}, they first generated the same number of adversarial test inputs as the number of original test inputs (also called natural test inputs), and then compared the test coverage of natural test inputs and that of both natural and adversarial test inputs.
They found that the latter is larger than the former, and thus concluded that the test coverage metrics are correlated with test effectiveness (measured by the generation of adversarial test inputs).
However, the experiments did not control for the number of test inputs when comparing test coverage, and thus the increased coverage may be caused by the generation of adversarial test inputs or just increasing the number of test inputs.

Further, we conducted a small experiment to investigate whether adding natural test inputs can also increase test coverage by taking the model \textit{LeNet-5} based on \textit{MNIST} and the test coverage \textit{KMNC} (to be introduced in Section~\ref{sec:structural}) as an example.
We first randomly selected 5,000 test inputs from \textit{MNIST} as the original test set and regarded the remaining 5,000 test inputs in \textit{MNIST} as the newly added natural test inputs, and then generated 5,000 adversarial test inputs via C\&W~\cite{DBLP:conf/sp/Carlini017} (to be introduced in Section~\ref{sec:advmethod}), a widely-used adversarial input generation method.
We found that the KMNC values of the original test set, the original test set and added natural test inputs, and the original test set and adversarial test inputs are 61.53\%, 68.77\%, and 65.87\%, respectively.
The results demonstrate that adding natural test inputs can also increase test coverage, even the increment is larger than that achieved by the same number of adversarial test inputs.
Hence, it is still unclear whether these test coverage metrics are correlated with test effectiveness by considering reasonable factors (e.g., controlling for the size of test sets).

Besides the experiments conducted in the work proposing the corresponding DNN test coverage metrics, an empirical study was also conducted to investigate the correlation between DNC~\cite{pei2017deepxplore} (the earliest DNN test coverage, to be introduced in Section~\ref{sec:structural}) and test effectiveness through using DNC to guide test input generation~\cite{harel2020neuron}.
They found increasing DNC is not a meaningful objective for generating DNN test inputs.
However, this study only investigated the earliest DNN test coverage metric, which cannot validate the fundamental assumption sufficiently.
Furthermore, there is another study investigating the usefulness of DNN test coverage from the perspective of model retraining quality (rather than the perspective of test effectiveness)~\cite{yan2020correlations}, and they found DNN test coverage does not have monotonic relations with model retraining quality.
Moreover, all these experiments and studies are conducted based on the subjects with limited diversity.
To sum up, the fundamental assumption for DNN test coverage is still not validated sufficiently and reasonably.
That is, \textit{the usefulness of DNN test coverage is still unclear from the test effectiveness perspective}.

To answer the open question, in this work we conducted an extensive study to revisit DNN test coverage from the test effectiveness perspective.
To sufficiently validate the fundamental assumption, we consider three test effectiveness criteria in DNN testing.
1) \textbf{Error-revealing capability}. Same as traditional software testing, revealing errors is also the most straightforward way of representing test effectiveness in DNN testing.
Here, we used the ratio of error-revealing test inputs in test sets to reflect the error-revealing capability of test sets, where error-revealing test inputs refer to those making the DNN produce incorrect predictions.
2) \textbf{Test diversity}. In DNN testing, besides paying attention to the overall accuracy of DNNs, it is also important to keep a watchful eye on \textit{each-class} accuracy for classification DNNs.
Here, we used the number of covered classes to represent test diversity, since it requires to ensure the existence of test inputs for testing each class before measuring each-class accuracy.
3) Subsequently, we have the third criterion, i.e., \textbf{error-revealing capability per diversity unit}, which is represented by the ratio of error-revealing test inputs in test sets that cover only a small number of classes (i.e., we regard 10\% of classes as a diversity unit in our study) rather than all classes (i.e., saturated diversity).
It can help reflect the diverse error-revealing capability in DNN testing.
In particular, to reasonably validate the fundamental assumption, we controlled the influence of the size of test sets when investigating the correlation between DNN test coverage and the three test effectiveness criteria.

In particular, we conducted a comprehensive study on DNN test coverage.
In our study, we carefully considered the diversity of subjects (10 pairs of models and datasets in total), involving different tasks, domains, DNN types, and model accuracy.
Besides the widely-used subjects in the existing studies~\cite{pei2017deepxplore,ma2018deepgauge,kim2019guiding,yan2020correlations,harel2020neuron,yang2022revisiting}, we also added five subjects that have never been used to evaluate any of the studied test coverage metrics before.
Besides, we not only studied all the DNN test coverage metrics used in the existing studies (except TKNP~\cite{ma2018deepgauge} since it is not a ratio but rather a number), but also added the state-of-the-art test coverage (i.e., IDC)~\cite{gerasimou2020importance}.

Different from the existing studies~\cite{yan2020correlations,harel2020neuron}, which delivered negative conclusions on the usefulness of existing DNN test coverage, our conclusions are relatively positive in general because we identified the practical usage scenarios for these existing DNN test coverage by investigating them from the test effectiveness perspective more sufficiently and reasonably.
Our study demonstrates that the studied structural coverage (i.e., DNC, TKNC, KMNC, and IDC) is complementary with the studied non-structural coverage (i.e., LSC and DSC), and they have different usage scenarios.
Specifically, the studied non-structural coverage (i.e., LSC and DSC~\cite{kim2019guiding}) has positive correlation with error-revealing capability regardless of saturated diversity or one diversity unit, but has no obvious positive correlation with test diversity.
Hence, they can be used to guide the generation of error-revealing test inputs but it is hard to guarantee the diversity of the generated test inputs guided by them.
The studied structural coverage (i.e., DNC, TKNC, KMNC, and IDC) does not have positive correlation with error-revealing capability under saturated diversity, but has positive correlation under small test diversity.
Hence, their usage scenario could be guiding the generation of error-revealing test inputs by taking a small number of classes of test inputs (rather than the whole test set with saturated diversity) as seed inputs\footnote{In existing DNN fuzzing and adversarial input generation methods, new test inputs are generated based on the given set of test inputs (also called seed inputs).}.
Also, they are positively correlated with test diversity,
and thus the guidance of these structural metrics can also help improve the diversity of generated test inputs.
More interesting findings and implications/suggestions can be found in Sections~\ref{results} and~\ref{sec:implications}.

To sum up, our work makes the following major contributions:
\vspace{-3mm}
\begin{itemize}
    \item We conducted a comprehensive study on DNN test coverage, which revisits the existing DNN test coverage metrics from the test effectiveness perspective. 
    
    \item We obtained some positive conclusions on the usefulness of existing DNN test coverage (which is different from all the existing studies) and identified the practical usage scenarios and promising research directions for them.
    
    \item We optimized and integrated all the implementations of collecting these test coverage metrics, and the coverage collection efficiency has been largely improved after our optimization (e.g., 89.83\%$\sim$95.54\% improvements on LeNet-5 and MNIST).   We have released our toolbox and all the experimental data on our project homepage~\cite{homepage}. 
\end{itemize}

\section{BACKGROUND}
\vspace{-1.5mm}
\label{sec:background}
In DNN testing, many test coverage metrics have been proposed to measure test effectiveness~\cite{pei2017deepxplore,ma2018deepgauge,kim2019guiding}. 
In our study, we not only studied all the test coverage metrics considered in the existing studies~\cite{yan2020correlations,harel2020neuron} (except TKNP~\cite{ma2018deepgauge} since it is not a ratio but rather a number) but also the recently-proposed test coverage (i.e., IDC~\cite{gerasimou2020importance}), in order to conduct a more comprehensive revisiting study.
According to whether considering structural elements are covered during testing, we classify test coverage metrics into two categories: \textit{structural coverage} and \textit{non-structural coverage}.
Here, we introduce these studied test coverage metrics in our work, and the details about other coverage will be presented in Section~\ref{sec:related}.



\vspace{-1.5mm}
\subsection{Structural Coverage}
\label{sec:structural}
Structural coverage considers which structural elements are covered during testing, mainly referring to various neuron coverage.
Here, we studied six typical structural coverage metrics.
The earliest one was proposed by Pei et al.~\cite{pei2017deepxplore}, which is called \textbf{DNC} (DeepXplore's Neuron Coverage).
DNC divides the state of a neuron into \textit{activated} and \textit{non-activated}, and measures the ratio of activated neurons in a DNN.
If the output value of a neuron is larger than a pre-defined threshold $t$ after executing a test input, DNC regards the neuron as an activated neuron.

Ma et al.~\cite{ma2018deepgauge} further proposed a set of neuron coverage metrics at several granularity levels, including TKNC (Top-K Neuron Coverage), KMNC (K-Multisection Neuron Coverage), NBC (Neuron Boundary Coverage), and SNAC (Strong Neuron Activation Coverage).
\textbf{TKNC} is a layer-level metric, which considers top-k neurons (ranked in the descending order of their output values after executing a test input) in a layer to be covered.
It measures the ratio of neurons that have once been top-k neurons in the corresponding layer.
\textbf{KMNC} is a more fine-grained metric, which first obtains the range of its output values for each neuron from training data (denoted as [$low_{i}$,$high_{i}$] for the neuron $n_{i}$) and then partitions the range into k sections.
If the output value of a neuron falls into a section after executing a test input, KMNC regards the section as a covered section.
KMNC measures the ratio of covered sections of all neurons in a DNN.
The insight behind KMNC is that, different sections may represent different functional modules of a DNN and more fine-grained neuron coverage can increase discrimination of different test inputs.
\textbf{NBC} also obtains the range of its output values for each neuron from training data like KMNC.
The difference is that, NBC considers whether the corner-case region, i.e., (${-\infty}$, ${low_i}$) and (${high_i}$, ${+\infty}$), of a neuron is covered after executing a test input.
\textbf{SNAC} is similar to NBC, but the difference is that SNAC only considers whether the upper corner-case region, i.e., (${high_i}$, ${+\infty}$), is covered after executing a test input.

Recently, Gerasimou et al.~\cite{gerasimou2020importance} proposed a novel test coverage metric \textbf{IDC} (Importance-Driven Coverage) that has never been investigated by the existing studies~\cite{yan2020correlations,harel2020neuron}. 
Its key insight is that more important neurons have a stronger causal relationship with other neurons and can explain which high-level features contribute more to decision-making, and thus the importance of each neuron is measured by the widely-used relevance score~\cite{bach2015pixel}.
Then, IDC partitions the space of each neuron’s activity into a set of clusters via neuron-wise quantization, and measures the ratio of combinations of important neurons clusters covered by a test set.

\subsection{Non-Structural Coverage}
\label{sec:nonstructural}
Non-structural coverage does not consider whether structural elements of a DNN are covered during testing.
The most typical non-structural coverage is surprise coverage~\cite{kim2019guiding}.
It regards the difference of DNN behaviors between a test input and training data as surprise adequacy (SA) of the test input.
Surprise coverage measures the range of SA values that a set of test inputs cover.
Specifically, Kim et al.~\cite{kim2019guiding} proposed two kinds of SA: Likelihood-based Surprise Adequacy (LSA) and Distance-based Surprise Adequacy (DSA).
LSA refers to the surprise of a test input with respect to the estimated density of each activation value in a set of activation traces observed over the neurons of a selected layer for training data via Kernel Density Estimation~\cite{wand1994kernel}.
LSA only considers the training data whose label is the same as the predicted class of the test input.
DSA is defined using the Euclidean distance between the activation trace observed over the neurons of a selected layer for a test input and a set of activation traces for training data.
DSA needs find the closest neighbor of the test input that shares the same class, denoted as $x_a$, and also the closest neighbor of $x_a$ in a different class.

According to LSA and DSA, the corresponding surprise coverage metrics are called \textbf{LSC}, and \textbf{DSC} respectively.
Specifically, given the upper and lower bounds of LSA/DSA denoted as \textit{U} and \textit{L}, LSC/DSC first divides [\textit{L},\textit{U}] into \textit{n} sections and then measures the ratio of covered sections.
If the LSA/DSA value of a test input falls into a section, LSC/DSC regards the section as a covered section.
In particular, a test input with a quite large SA value may be irrelevant to the problem domain (e.g., an image of an apple is irrelevant to the testing of traffic sign classifiers), and thus the parameter \textit{U} can be used to filter out those irrelevant test inputs.
Please note that DSC is not applicable to regression models~\cite{kim2019guiding}.


\vspace{-3mm}
\section{STUDY DESIGN}
\vspace{-3mm}
Our study aims to revisit the existing DNN test coverage metrics from the test effectiveness perspective, thus answering the open question: \textit{are these existing metrics correlated with test effectiveness}. 
As presented in Section~\ref{sec:intro}, we consider three test effectiveness criteria (i.e., error-revealing capability 
, test diversity, and error-revealing capability per diversity unit) in DNN testing to sufficiently validate the fundamental assumption.
Specifically, we aim to answer it by addressing the following research questions (RQs): 

\begin{itemize}
\item \textbf{RQ1:} Are the studied test coverage metrics correlated with the error-revealing capability of test sets (measured by the ratio of error-revealing test inputs in test sets)?
\item \textbf{RQ2:} Are the studied test coverage metrics correlated with the test diversity of test sets (measured by the ratio of covered classes in test sets)?
\item \textbf{RQ3:} Are the studied test coverage metrics correlated with error-revealing capability per diversity unit?
\end{itemize}
Please note that for a classification model, we treat covering all classes in a test set as \textit{saturated diversity} and covering 10\% of classes as \textit{a diversity unit}.
That is, RQ3 aims to investigate the correlation between test coverage and the error-revealing capability of test sets that covers 10\% of classes.
Actually, RQ1 investigates the correlation under saturated diversity and
RQ3 is inspired by the conclusions from RQ1 and RQ2.

\subsection{Datasets and DNN Models}

\begin{table}[bt]
\caption{\label{tab:subs} Datasets and DNN models}
\begin{adjustbox}{max width=1.0\textwidth,center}
\begin{threeparttable}
\begin{tabular}{lllrrrrrrrr}
\headrow
\thead{ID}&\thead{Dataset}&\thead{Model}&\thead{Size(KB)}&\thead{\#Train}&\thead{\#Test}&\thead{Acc(\%)}&\thead{Task}&\thead{Domain}&\thead{Network}&\thead{Used}\\
1&MNIST&LeNet-5&1,300&60,000&10,000&98.72& classification&image&CNN&\yes{}\\
2&Fashion-MNIST&LeNet-5&1,279&60,000&10,000&91.07& classification&image&CNN&\no{}\\
3&CIFAR-10&AlexNet&9,830&50,000&10,000&76.64& classification&image&CNN&\no{}\\
4&CIFAR-10&ResNet-20&3,507&50,000&10,000&91.21& classification&image&CNN&\yes{}\\
5&CIFAR-10&VGG-16&19,639&50,000&10,000&87.41& classification&image&CNN&\yes{}\\
6&CIFAR-100&ResNet-32&10,615&50,000&10,000&70.52& classification&image&CNN&\no{}\\
7&Driving&Dave-orig&8,306&101,396&5,614&90.33&regression&image&CNN&\yes{}\\
8&Driving&Dave-dropout&13,139&101,396&5,614&91.82&regression&image&CNN&\yes{}\\
9&Speech-Commands&DeepSpeech& 6,734& 51,776 & 6,471 &94.53& classification&speech&RNN&\no{}\\
10& 20-Newsgroups &TextCNN&38,974&11,314&7,532&77.68&classification&text&CNN&\no{}\\
\hline  
\end{tabular}

\begin{tablenotes}
\small
\item[*] We use 1 - MSE (Mean Square Error) as the accuracy of regression models.
\yes{} represents the subject has been used to evaluate at least one of the studied test coverage metrics before. 
\no{} represents the subject has never been used to evaluate any of the studied test coverage metrics before.
\end{tablenotes}
\end{threeparttable}
\end{adjustbox}
\vspace{-5mm}
\end{table}

In our study, we used 10 pairs of datasets and DNN models as subjects in total. 
Table~\ref{tab:subs} shows the basic information on subjects, where Columns 4-10 represent the model size, the number of training instances, the number of original test inputs, the model accuracy, the model task, the domain of the model, and the network type, respectively. 
Specifically, MNIST is a handwritten digit dataset~\cite{mnist}, Fashion-MNIST is a MNIST-like dataset with greyscale images of Zalando's items~\cite{xiao2017/online}.
CIFAR-10 and CIFAR-100 are 10-class and 100-class ubiquitous object datasets respectively~\cite{cifar}, 
Driving is an Udacity-provided autonomous driving dataset~\cite{driving}, and Speech-Commands is a sequential dataset containing a set of one-second .wav audio files, each of which contains a single spoken English word~\cite{speech}. 20-Newsgroups is a text dataset containing about 20,000 newsgroup documents, which has been widely used in natural language processing (NLP) tasks (e.g., text classification~\cite{20news}).

To more comprehensively study these test coverage metrics, \textit{we carefully considered the diversity of subjects}, including 1) \textbf{different tasks}: normal classification, multi-label classification (DeepSpeech), and regression; 2) \textbf{different domains}: image domains, speech domains, and text domains; 3) \textbf{different DNN types}: CNN and RNN; and 4) \textbf{different model accuracy}: from 70.52\% to 98.72\%. 

In our study, we not only selected five subjects that have been used to evaluate at least one of the studied test coverage metrics before without any subjective bias, but also used \textbf{five subjects that have never been used to evaluate any of the studied test coverage metrics before}, which is shown in the last column in Table~\ref{tab:subs}.



\subsection{Implementations}
\label{sec:imple}


Since the code for collecting DNC, LSC, DSC, and IDC is released by the existing work~\cite{pei2017deepxplore,kim2019guiding,gerasimou2020importance}, we directly adapted the code to collect them in our study.
As for TKNC, KMNC, NBC, and SNAC, we communicated with the corresponding authors and they shared the implementations with us.
After carefully reviewing the implementations, we discussed with some authors about potential problems in code, and further optimized the code for KMNC, TKNC, NBC, SNAC, and IDC in order to improve their coverage collection efficiency.
Table~\ref{tab:effiency} shows the time spent on collecting these coverage metrics when executing 10,000 test inputs of the subject (ID: 1) before and after our optimization. 
Column \textit{Original} presents the time spent using the original implementation without our optimization, while Column \textit{Optimized} presents the time spent using our optimized implementation. 
The last column shows the efficiency improvement of our optimized implementation over the original one. 
We found that the coverage collection efficiency is largely improved, i.e., 89.83\%$\sim$95.54\% improvements.
In particular, we checked that the coverage results are the same before and after our optimization.


\begin{wraptable}{l}{6cm}
    \caption{Coverage collection efficiency comparison between before and after our optimization for TKNC, KMNC, NBC, SNAC, and IDC on the subject (ID: 1) (seconds)}\label{tab:effiency} 
    \begin{adjustbox}{max width=0.9\textwidth,left}
    \begin{threeparttable}
    \begin{tabular}{lrrl}
    \headrow
    \thead{Coverage} &  \thead{Original} & \thead{Optimized} &  \thead{$\Uparrow_{\textit{rate}}$(\%)} \\
    TKNC & 40.10 & 3.40 & 91.52\% \\
    KMNC & 76.30 & 3.40 & 95.54\% \\
    NBC & 59.10 & 3.40 & 94.25\% \\
    SNAC & 57.20 & 3.30 & 94.23\% \\
    IDC & 4,465.80 & 454.00 & 89.83\% \\
    \hline  
    \end{tabular}
    \end{threeparttable}
    \end{adjustbox}
    \vspace{-4mm}
\end{wraptable}


To promote future research and practical usage, we have developed a toolbox that integrates the implementations of collecting the eight test coverage, each of which can be invoked by simply specifying the corresponding parameter value.
Our toolbox and all the experimental data have been released on our project homepage~\cite{homepage}



For the parameters of these studied coverage metrics, we set them following the existing work~\cite{kim2019guiding,pei2017deepxplore,ma2018deepgauge,gerasimou2020importance}.
The $t$ value of DNC is set to 0.5, the $k$ value of TKNC is set to 10, the $k$ value of KMNC is set to 1,000, and the $n$ value of LSC and DSC is set to 1,000.
LSC and DSC also have some parameters required to be specifically set for different subjects~\cite{kim2019guiding}, i.e., the selected layer to calculate LSA/DSA, the upper bound $U$ and lower bound $L$ of LSC/DSC.
However, there is no guide provided to help set them for different subjects.
Hence, by communicating with the authors and referring to their implementation, for each subject we set the upper bound $U$ and the lower bound $L$ of LSC/DSC to the largest and smallest LSA/DSA values of all the test inputs respectively.
We further investigated the influence of $U$ on the performance of LSC/DSC in Section~\ref{rq1_non_structural}.
Also, we set the selected layer required by LSA/DSA to the last-hidden layer of a model according to the existing work~\cite{Li:2019:BOD:3338906.3338930,chenpracticalTOSEM}, since the work has demonstrated that deeper layers tend to perform better in DNN testing.

We conducted our experiments on the Intel Xeon Silver 4214 machine with 128GB RAM, Ubuntu 18.04 LTS, and eight GTX 2080 Ti GPUs.
We used the Python interpreter with version 3.6.13 and utilized the Anaconda to manage Python packages. 

\subsection{Experimental Setup}
\label{sec:setup}
In this section, we present the experimental setup to answer each RQ for each subject.
Please note that we used all the subjects in RQ1, while used all the normal classification models (ID: 1$\sim$6)
in both RQ2 and RQ3 since regression models do not have the concept of ``class'' and the number of 10\% of classes of test inputs in the subjects (ID: 9 and ID: 10) is too small to support the setups of RQ2 and RQ3.


\subsubsection{RQ1's Setup}
\label{rq1} 
RQ1 aims to investigate the correlation between test coverage and error-revealing capability (under saturated diversity).
We used the ratio of error-revealing test inputs in a test set to measure the error-revealing capability of the test set.
To answer RQ1, we constructed $m$ test sets with various ratios of error-revealing test inputs. 
In particular, we controlled for the size of test sets, i.e., creating the test sets with the same size $n$.
In practice, each test set should contain a number of passing test inputs (that make the DNN produce correct prediction) since passing test inputs are common for a well-trained DNN.
Hence, we ensure that the percentage  of error-revealing test inputs in each test set is less than or equal to $\alpha$\%.
That is, when constructing a test set, we first randomly determined the percentage of error-revealing test inputs $r$\% from [0, $\alpha\%$] (i.e., 0$\leq$$r$\%$\leq$$\alpha\%$) and randomly sampled $r\%$*$n$ error-revealing test inputs. Then, we randomly sampled the remaining (1-$r\%$)*$n$ passing test inputs from the whole set of natural test inputs to construct the test set.
Here, we set $m=500$, $n=800$, $\alpha\%=70\%$. 
The whole set of test inputs refers to all the natural test inputs provided by the original dataset and the generated adversarial test inputs.
This is because the number of \textit{natural} error-revealing test inputs tends to be small in the original set of test inputs, we then generated adversarial test inputs as supplementary to increase the number of error-revealing test inputs.
We will introduce the adversarial test input generation methods used in our study in Section~\ref{sec:advmethod}.
Please note that we ensure that each test set for classification models has saturated diversity.

Further, we collected the studied eight DNN test coverage for each test set, and then measured the correlation between each test coverage and error-revealing capability (i.e., the ratio of error-revealing test inputs in test sets).
In our study, we adopted the widely-used \textbf{Spearman correlation} method~\cite{myers2004spearman}, which is used to assess how well the relationship between two variables can be described using a \textit{monotonic} function.
The sign of the Spearman correlation coefficient (denoted as $c$) represents the positive or negative correlation between two variables, $|c|$ represents the correlation degree, and the $p$ value reported by the Spearman correlation method represents whether the correlation is statistically significant (at the level of 0.05).
According to the existing work~\cite{inozemtseva2014coverage,chen2017assertions}, the correlation can be divided into four categories: weak correlation (0<$|c|\leq$0.4), moderate correlation (0.4<$|c|\leq$0.7), strong correlation (0.7<$|c|\leq$0.9), and extremely strong correlation (0.9<$|c|\leq$1).

\begin{figure*}[t!]
\centering
\subfloat[The investigation on the influence of parameter $m$, studying the value of $m$ ranging from 100 to 1000 with the interval of 100.]{
\hspace{-5mm}
\includegraphics[scale=0.40]{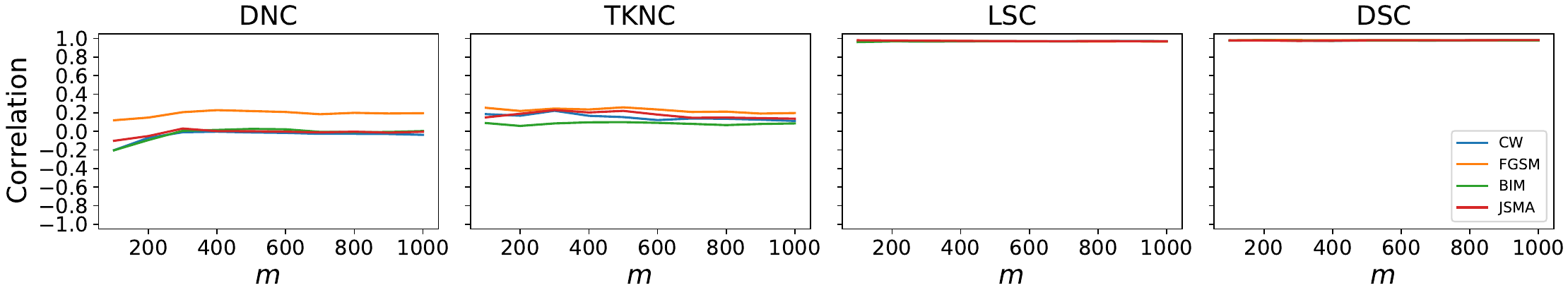}
}
\vspace{-2mm}
\subfloat[The investigation on the influence of parameter $n$, studying the value of $n$ ranging from 500 to 1000 with the interval of 100. ]{
\hspace{-5mm}
\includegraphics[scale=0.40]{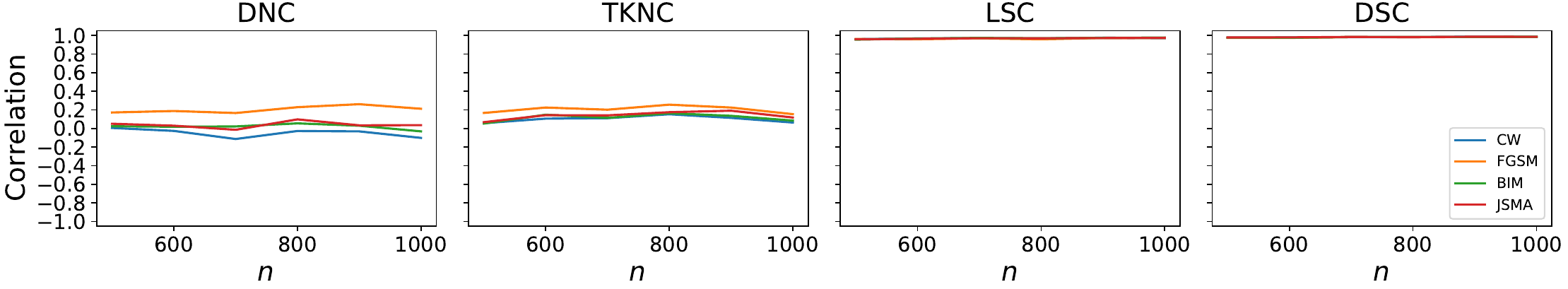}
}
\vspace{-2mm}
\subfloat[The investigation on the influence of parameter $\alpha$, studying the value of $\alpha$ ranging from 0.5 to 0.9 with the interval of 0.1.]{
\hspace{-5mm}
\includegraphics[scale=0.40]{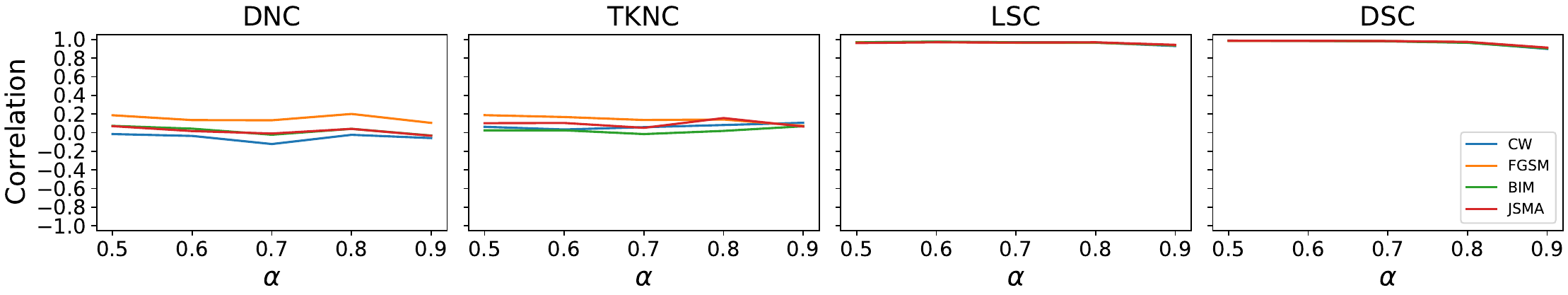}
}
\caption{The influence of $m$, $n$ and $\alpha$ on the correlation coefficient between test coverage and error-revealing capability on the subject (ID: 1), respectively.}
\label{fig:parameter}
\end{figure*}

For the parameters of  $m$, $n$, and $\alpha$ (afftecting r), we have conducted experiments to investigate the influence of these parameters by using MNIST-LeNet5 (ID:1) as the representative. Here, we used DNC and TKNC as the representative of structural coverage and also studied the two non-structural coverage (LSC and DSC). We studied the value of $m$ ranging from 100 to 1000 with the interval of 100, the value of $n$ ranging from 500 to 1000 with the interval of 100, and the value of $\alpha$ ranging from 0.5 to 0.9 with the interval of 0.1, respectively. The results are shown in the Figure~\ref{fig:parameter}, where the x-axis represents the settings of the studied parameter while the y-axis represents the Spearman correlation coefficient between test coverage and error-revealing capability. From the figure, we conclude that the influence of the parameter $m$ under the studied settings is very small, indicating the generality of our conclusions to some degree.


\subsection{RQ2: Correlation between Test Coverage and Test Diversity}

\subsubsection{RQ2's Setup}
\label{rq2}
RQ2 aims to investigate the correlation between test coverage and test diversity.
We used the ratio of covered classes (among all classes involved in the classification model) in a test set to measure the test diversity of the test set.
To answer RQ2, we constructed a number of test sets with various ratios of covered classes. 
In particular, we controlled for both the size of test sets (denoted as $n$) and the ratio of error-revealing test inputs in test sets (denoted as $r\%$).
As presented before, we treat covering 10\% of classes in a test set as a diversity unit. Regarding the chosen 10\%, we set it based on the characteristics of our subjects. Since the subjects used in our study contain at least 10 classes for classification models, ``10\%'' is the smallest unified setting ensuring that the diversity unit for each subject can contain at least one class. Moreover, ``10\%'' can be a good trade-off between effectiveness (having statistical significance) and efficiency (having the reasonable number of test sets for coverage collection) in our experiments.

We first evenly divided the whole set of test inputs into 10 groups according to the ascending order of the class IDs.
By taking CIFAR-100 (including 100 classes) as an example, we put all the natural test inputs with the class IDs $0\sim 9$ and the generated adversarial test inputs based on these natural test inputs into the first group.
Then, from the first $i$ ($i\in\{1,2,\ldots,10\}$) groups, we constructed $m$ test sets, each of which is formed by randomly sampling $r\%$*$n$ error-revealing test inputs and (1-$r\%$)*$n$ passing test inputs.
Correspondingly, the test diversity of each test set constructed from the first $i$ groups is $i$ diversity units. 
Indeed, we ensure that each test set has the corresponding numbers of diversity units.
In RQ2, we set $n=800$, $r\%=50\%$, $m=100$.
Next, we collected the eight test coverage for each test set, and then measured the Spearman correlation between each test coverage and test diversity.



\subsubsection{RQ3's Setup}
\label{rq3} 

By comprehensively considering RQ1 and RQ2, RQ3 subsequently investigates the correlation between test coverage and error-revealing capability per diversity unit.
Actually, RQ1 considers all classes (i.e., saturated diversity) to investigate the correlation.
RQ3 repeats the experiment of RQ1 under one diversity unit.
We constructed $m$ test sets ($m=100$), and when constructing a test set, we sampled test inputs from \textit{one} group of test inputs (defined in Section~\ref{rq2}) instead of the whole set of test inputs in RQ1.
The remaining settings are the same as RQ1.
To reduce the bias from the selection of the target group, we repeated this experiment \textit{five} times by randomly selecting five different groups as the target groups for test set construction respectively.
For each time, we collected the eight test coverage for each test set, and then measured the correlation between each test coverage and error-revealing capability per diversity unit via Spearman correlation.

Besides, we investigated the trend of the correlation between test coverage and error-revealing capability with the test diversity increasing from one diversity unit to saturated diversity.
We additionally considered $i$ ($i\in\{2,\ldots, 9\}$) diversity units by constructing test sets from the first $i$ groups of test inputs respectively.
That is, we repeated the above one-diversity-unit (or saturated-diversity) experiment under $i$ ($i\in\{2,\ldots, 9\}$) diversity units respectively.

\subsection{Adversarial Test Input Generation Methods}
\label{sec:advmethod}
In our study, we adopted four advanced adversarial test input generation methods to generate adversarial test inputs for general image datasets, i.e., MNIST, Fashion-MNIST, CIFAR-10, and CIFAR-100.
They are \textit{FGSM} (Fast Gradient Sign Method)~\cite{DBLP:journals/corr/GoodfellowSS14}, \textit{C\&W} (Carlini\&Wagner)~\cite{DBLP:conf/sp/Carlini017}, \textit{BIM} (Basic Iterative Methods)~\cite{DBLP:conf/iclr/KurakinGB17a}, and \textit{JSMA} (Jacobian-based Saliency Map Attack)~\cite{DBLP:conf/eurosp/PapernotMJFCS16}.
These methods have been widely used in the existing work~\cite{ma2018deepgauge,kim2019guiding}.
For the dataset Driving, we used the two methods proposed by DeepXplore~\cite{pei2017deepxplore} (called \textit{Patch} and \textit{Light} in our work)
to generate adversarial test inputs following most of the existing work using this dataset~\cite{kim2019guiding,Li:2019:BOD:3338906.3338930,chenpracticalTOSEM}.
For the datasets Speech-Commands and 20-Newsgroups, all the above-mentioned methods cannot be applicable to them since they are not images, but audios and texts respectively.
Therefore, we adopted the widely-used \textit{CTCM} (CTC loss based Method)~\cite{DBLP:conf/sp/Carlini018} and \textit{PWWS} (Probability Weighted Word Saliency)~\cite{pwws2019} to generate adversarial test inputs for them, respectively.
\section{Results and Analysis}\label{results}
\vspace{-1.5mm}
We adopt the following representation paradigm for tables in this paper:
The \textbf{bold} value represents that the $p$ value is smaller than 0.05, indicating the correlation has statistical significance. The value marked with \legendsquare{MyGreen!15!white} represents \textbf{moderate} positive correlation, \legendsquare{MyGreen!50!white} represents \textbf{strong} positive correlation and \legendsquare{MyGreen!100!white} represents \textbf{extremely strong} positive correlation. The cell marked with ``---'' represents the coverage is not applicable to the corresponding model.
\vspace{-3mm}
\subsection{RQ1: Correlation between Test Coverage and Error-revealing Capability}
\subsubsection{Results on Structural Coverage}
\begin{wraptable}[21]{l}{0.47\textwidth}
    \vspace{-5mm}
    \caption{\label{tab:rq1cor} RQ1: Spearman correlation between test coverage and the ratio of error-revealing test inputs.}
    
    \begin{adjustbox}{max width=0.5\textwidth,left}
    \begin{threeparttable}
    \begin{tabular}{clrrrrrrrr}
    \headrow
    \thead{ID} & \thead{Adv.} & \thead{DNC} & \thead{NBC} & \thead{SNAC} & \thead{TKNC} & \thead{KMNC} & \thead{IDC} & \thead{LSC} & \thead{DSC} \\
    \hiderowcolors
    \multirow{4}{*}{1} & CW & -0.05 & \textbf{-0.3} & \textbf{-0.37} & \textbf{0.11} & \textbf{-0.94} & 0.08 & \cellcolor{MyGreen!100!black}\textbf{0.97} & \cellcolor{MyGreen!100!black}\textbf{0.98} \\
     & FGSM & \textbf{0.22} & \cellcolor{MyGreen!15!white}\textbf{0.51} & \textbf{0.3} & \textbf{0.2} & \textbf{-0.97} & \textbf{-0.53} & \cellcolor{MyGreen!100!black}\textbf{0.97} & \cellcolor{MyGreen!100!black}\textbf{0.98} \\
     & BIM & \textbf{0.09} & \textbf{-0.25} & \textbf{-0.34} & 0.09 & \textbf{-0.94} & -0.03 & \cellcolor{MyGreen!100!black}\textbf{0.97} & \cellcolor{MyGreen!100!black}\textbf{0.98} \\
     & JSMA & 0.08 & \textbf{0.3} & \textbf{-0.32} & \textbf{0.2} & \textbf{-0.85} & 0.0 & \cellcolor{MyGreen!100!black}\textbf{0.97} & \cellcolor{MyGreen!100!black}\textbf{0.98} \\ \hline
    \multirow{4}{*}{2} & CW & 0.08 & \textbf{-0.09} & \textbf{-0.14} & -0.02 & \textbf{-0.82} & \textbf{0.32} & \cellcolor{MyGreen!100!black}\textbf{0.97} & \cellcolor{MyGreen!100!black}\textbf{0.95} \\
     & FGSM & \cellcolor{MyGreen!15!white}\textbf{0.4} & \cellcolor{MyGreen!15!white}\textbf{0.59} & \cellcolor{MyGreen!15!white}\textbf{0.57} & 0.03 & \textbf{-0.21} & -0.02 & \cellcolor{MyGreen!100!black}\textbf{0.97} & \cellcolor{MyGreen!100!black}\textbf{0.97} \\
     & BIM & \textbf{0.19} & \cellcolor{MyGreen!15!white}\textbf{0.54} & \cellcolor{MyGreen!15!white}\textbf{0.54} & -0.0 & \textbf{-0.73} & \textbf{0.38} & \cellcolor{MyGreen!100!black}\textbf{0.97} & \cellcolor{MyGreen!100!black}\textbf{0.96} \\
     & JSMA & \textbf{0.09} & 0.01 & \textbf{-0.14} & 0.01 & \textbf{-0.71} & \textbf{0.39} & \cellcolor{MyGreen!100!black}\textbf{0.98} & \cellcolor{MyGreen!100!black}\textbf{0.96} \\\hline
    \multirow{4}{*}{3} & CW & 0.01 & \textbf{-0.34} & \textbf{-0.34} & 0.02 & \textbf{-0.96} & \textbf{-0.66} & \cellcolor{MyGreen!100!black}\textbf{0.95} & \cellcolor{MyGreen!50!white}\textbf{0.75} \\
     & FGSM & \textbf{0.25} & \cellcolor{MyGreen!50!white}\textbf{0.8} & \cellcolor{MyGreen!50!white}\textbf{0.79} & 0.08 & -0.02 & \textbf{-0.37} & \cellcolor{MyGreen!100!black}\textbf{0.94} & \cellcolor{MyGreen!100!black}\textbf{0.92} \\
     & BIM & 0.08 & \cellcolor{MyGreen!15!white}\textbf{0.59} & \cellcolor{MyGreen!15!white}\textbf{0.59} & 0.05 & \textbf{-0.92} & \textbf{-0.5} & \cellcolor{MyGreen!100!black}\textbf{0.94} & \cellcolor{MyGreen!50!white}\textbf{0.85} \\
     & JSMA & -0.04 & \textbf{-0.3} & \textbf{-0.31} & 0.04 & \textbf{-0.98} & \textbf{-0.77} & \cellcolor{MyGreen!100!black}\textbf{0.94} & \cellcolor{MyGreen!50!white}\textbf{0.81} \\ \hline
    \multirow{4}{*}{4} & CW & -0.03 & -0.04 & -0.05 & 0.0 & \textbf{-0.75} & \cellcolor{MyGreen!15!white}\textbf{0.45} & \cellcolor{MyGreen!50!white}\textbf{0.87} & \textbf{0.36} \\
     & FGSM & \textbf{0.12} & \cellcolor{MyGreen!15!white}\textbf{0.59} & \cellcolor{MyGreen!15!white}\textbf{0.62} & \cellcolor{MyGreen!15!white}\textbf{0.44} & \textbf{-0.54} & \textbf{0.24} & \cellcolor{MyGreen!50!white}\textbf{0.8} & \cellcolor{MyGreen!15!white}\textbf{0.59} \\
     & BIM & -0.05 & -0.06 & -0.04 & -0.04 & \textbf{-0.75} & \textbf{0.37} & \cellcolor{MyGreen!50!white}\textbf{0.86} & \cellcolor{MyGreen!15!white}\textbf{0.4} \\
     & JSMA & \textbf{-0.11} & \textbf{-0.11} & -0.07 & \textbf{-0.17} & \textbf{-0.87} & \textbf{0.27} & \cellcolor{MyGreen!50!white}\textbf{0.84} & \textbf{0.32} \\ \hline
    \multirow{4}{*}{5} & CW & \textbf{-0.58} & \textbf{-0.22} & \textbf{-0.22} & \textbf{0.38} & \textbf{-0.95} & \textbf{-0.71} & \cellcolor{MyGreen!15!white}\textbf{0.58} & \cellcolor{MyGreen!100!black}\textbf{0.92} \\
     & FGSM & \textbf{-0.63} & \cellcolor{MyGreen!15!white}\textbf{0.47} & \cellcolor{MyGreen!15!white}\textbf{0.47} & \textbf{0.32} & \textbf{-0.98} & \textbf{-0.77} & \textbf{0.18} & \cellcolor{MyGreen!100!black}\textbf{0.95} \\
     & BIM & \textbf{-0.64} & \cellcolor{MyGreen!15!white}\textbf{0.42} & \cellcolor{MyGreen!15!white}\textbf{0.42} & \textbf{0.33} & \textbf{-0.97} & \textbf{-0.77} & \textbf{0.24} & \cellcolor{MyGreen!100!black}\textbf{0.94} \\
     & JSMA & \textbf{-0.64} & \textbf{-0.27} & \textbf{-0.26} & \textbf{0.3} & \textbf{-0.97} & \textbf{-0.78} & \textbf{0.26} & \cellcolor{MyGreen!100!black}\textbf{0.94} \\ \hline
    \multirow{4}{*}{6} & CW & \textbf{-0.16} & \textbf{-0.13} & \textbf{-0.11} & \textbf{-0.33} & \textbf{-0.51} & \textbf{-0.19} & \cellcolor{MyGreen!15!white}\textbf{0.51} & \textbf{-0.49} \\
     & FGSM & \textbf{0.22} & \cellcolor{MyGreen!15!white}\textbf{0.6} & \cellcolor{MyGreen!15!white}\textbf{0.61} & \textbf{-0.13} & \textbf{0.15} & \textbf{-0.2} & \cellcolor{MyGreen!50!white}\textbf{0.81} & \textbf{-0.26} \\
     & BIM & \textbf{-0.11} & -0.06 & -0.04 & \textbf{-0.28} & \textbf{-0.49} & \textbf{-0.14} & \cellcolor{MyGreen!15!white}\textbf{0.58} & \textbf{-0.5} \\
     & JSMA & \textbf{-0.17} & \textbf{-0.15} & \textbf{-0.16} & \textbf{-0.5} & \textbf{-0.68} & -0.05 & \cellcolor{MyGreen!15!white}\textbf{0.54} & \textbf{-0.6} \\ \hline

    \multirow{2}{*}{7} & PATCH & \cellcolor{MyGreen!15!white}\textbf{0.65} & \cellcolor{MyGreen!50!white}\textbf{0.77} & \cellcolor{MyGreen!50!white}\textbf{0.77} & \cellcolor{MyGreen!50!white}\textbf{0.7} & \cellcolor{MyGreen!50!white}\textbf{0.81} & \cellcolor{MyGreen!50!white}\textbf{0.71} & \cellcolor{MyGreen!15!white}\textbf{0.49} & --- \\
     & LIGHT & \textbf{-0.76} & \cellcolor{MyGreen!50!white}\textbf{0.86} & \cellcolor{MyGreen!50!white}\textbf{0.86} & \textbf{-0.8} & \cellcolor{MyGreen!50!white}\textbf{0.88} & \textbf{-0.4} & \cellcolor{MyGreen!50!white}\textbf{0.87} & --- \\ \hline
    \multirow{2}{*}{8} & PATCH & \textbf{0.39} & \cellcolor{MyGreen!50!white}\textbf{0.79} & \cellcolor{MyGreen!50!white}\textbf{0.79} & \textbf{0.2} & \cellcolor{MyGreen!50!white}\textbf{0.78} & \cellcolor{MyGreen!15!white}\textbf{0.65} & \cellcolor{MyGreen!50!white}\textbf{0.75} & --- \\ 
     & LIGHT & \cellcolor{MyGreen!50!white}\textbf{0.8} & \textbf{-0.65} & \textbf{-0.65} & \cellcolor{MyGreen!50!white}\textbf{0.73} & 0.08 & \textbf{-0.13} & \cellcolor{MyGreen!50!white}\textbf{0.89} & --- \\ \hline
    9 & CTCM & \textbf{-0.68} & \textbf{0.27} & \textbf{0.34} & \textbf{-0.91} & \textbf{-0.79} & --- & \cellcolor{MyGreen!50!white}\textbf{0.83} & \cellcolor{MyGreen!100!black}\textbf{0.96} \\ \hline
    10 & PWWS & \textbf{-0.15} & \textbf{-0.2} & \textbf{-0.23} & \textbf{-0.12} & \cellcolor{MyGreen!50!white}\textbf{0.8} & \cellcolor{MyGreen!50!white}\textbf{0.84} & \cellcolor{MyGreen!100!black}\textbf{0.96} & \textbf{0.21}\\
    
    \hline  
    \end{tabular}
    \end{threeparttable}
    \end{adjustbox}
\end{wraptable}
Table~\ref{tab:rq1cor} shows the Spearman correlation results between test coverage and error-revealing capability (under saturated diversity). From Columns 3-8, in general, the six structural coverage metrics have moderate to extremely strong positive correlation with the error-revealing capability of test sets (i.e., the ratio of error-revealing test inputs in test sets) \textit{in few cases}, but have weak positive correlation, even no significant correlation and negative correlation, \textit{in most cases}.
For example, TKNC has weak positive correlation in 26.67\% cases, no significant correlation in 36.67\% cases, and negative correlation in 26.67\% cases, with the ratio of error-revealing test inputs, but has moderate to extremely strong positive correlation in only 10\% cases.

This conclusion is similar to the one obtained in the existing study on DNC (i.e., increasing DNC is not a meaningful objective in DNN testing)~\cite{harel2020neuron}.
That indicates when controlling for the size and saturated diversity of each test set, both passing test inputs and error-revealing test inputs have the similar capability to increase these structural coverage metrics, and sometimes the coverage-increasing capability of the former is stronger than that of the latter.

Through further observation, the results on different structural coverage have some differences.
Among the cases with moderate to extremely strong positive correlation for ID:1$\sim$6, 55\% of them occur on NBC and SNAC when adopting FGSM to generate adversarial test inputs.
The reason is that when generating adversarial test inputs, C\&W, BIM, and JSMA aim to minimize the difference between an original test input and an adversarial test input or perform a tiny perturbation in a single step while FGSM does not consider them.
Hence, FGSM is more likely to generate the adversarial test inputs that have farther distance from the natural test inputs, further leading to irrelevant test inputs (that are meaningless in DNN testing). 
Those more different, even irrelevant test inputs, are much easier to cover corner-case regions of the output values for neurons, leading to stronger positive correlation for NBC and SNAC.
However, such positive correlation may be meaningless due to irrelevant test inputs.
That also indicates in the studies of DNN testing, it is not enough to only adopt FGSM to generate adversarial test inputs due to their different characteristics.
Besides, almost all the cases for KMNC have negative correlation.
We carefully analyzed the reason and found that with the ratio of error-revealing test inputs increasing, the sections closing to the boundary $low_{i}$ and $high_{i}$ are covered more frequently, indicating that more test inputs focus on covering a small number of sections.
This leads to the decrement of the overall KMNC coverage.



In general, the correlation on regression models is stronger than classification models.
This is because the accuracy of regression models is measured based on MSE, which is more fine-grained, and thus is more sensitive to capture differences in test behaviors (such as neuron coverage differences).
That suggests if we can design a more fine-grained accuracy measurement for classification models, the correlation between test coverage and error-revealing capability may be effectively improved.

\subsubsection{Results on Non-structural Coverage}\label{rq1_non_structural}

\begin{wrapfigure}[11]{l}{0.6\textwidth}
 \center
 \vspace{-3mm}
  \includegraphics[scale=0.35]{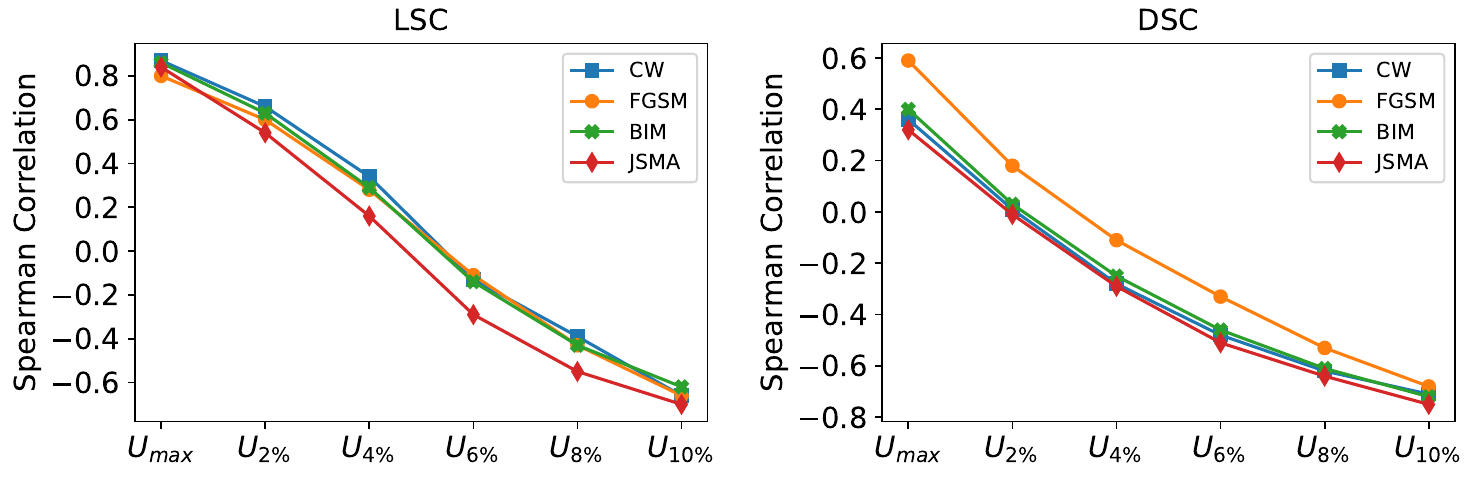}
 \vspace{-5mm}
 \caption{\label{fig:parameter influence} Spearman correlation under different settings of $U$ for LSC and DSC on the subject (ID: 4).}
 \vspace{-5mm}
\end{wrapfigure}
We then analyzed the correlation between non-structural coverage and error-revealing capability from the last two columns in Table~\ref{tab:rq1cor}.
Different from structural coverage, both LSC and DSC have moderate to extremely strong positive correlations \textit{in almost all the cases}.
The results demonstrate that measuring the difference of DNN behaviors between test inputs and training data is effective to measure the error-revealing capability of test sets, when controlling for the size and the saturated diversity of test sets.
That is, the non-structural coverage outperforms the structural coverage in this scenario, and LSC slightly outperforms DSC.

But the remarkable thing for LSC and DSC is that some parameters in them are required to be specially set for different subjects.
These parameters may affect the performance of LSC/DSC, but there is no guide to help set them for different subjects.
Here, we conducted an experiment to investigate the influence of $U$ on both LSC and DSC since this parameter is very important, which is used to filter out irrelevant test inputs.
We studied six settings of $U$: $U_{max}$ (the maximum value of LSA/DSA), which is the setting used in our study and means that we did not filter out any test inputs (even though some are irrelevant), and $U_{2\%}\sim U_{10\%}$ with the step size of $2\%$ (the 2\%$\sim$10\% quantile values of all the ranked LSA/DSA values in their descending order), which simulates different degrees of filtering out irrelevant test inputs.

Figure~\ref{fig:parameter influence} shows the results, where the x-axis represents different settings of $U$ while the y-axis represents the Spearman correlation coefficient between LSC/DSC and error-revealing capability.
From this figure, the correlation decreases monotonically as the upper bound $U$ becomes smaller, demonstrating the significant influence of $U$. 
This is because more test inputs with large LSA/DSA values are filtered out, which are more likely to contain both irrelevant test inputs and really error-revealing test inputs (especially when U is not set properly). This demonstrates the large influence of the parameter U and the importance of providing the guide to set the parameter U properly.

Therefore, providing a guide for setting the parameters of LSC and DSC or designing an automatic method of adaptively setting the parameters for different subjects is an important future work for improving LSC and DSC.
\vspace{-5mm}

\begin{figure*}[t!]
 \center

 \begin{tabular}{l}
   \hspace{-5mm}
  \includegraphics[scale=0.30]{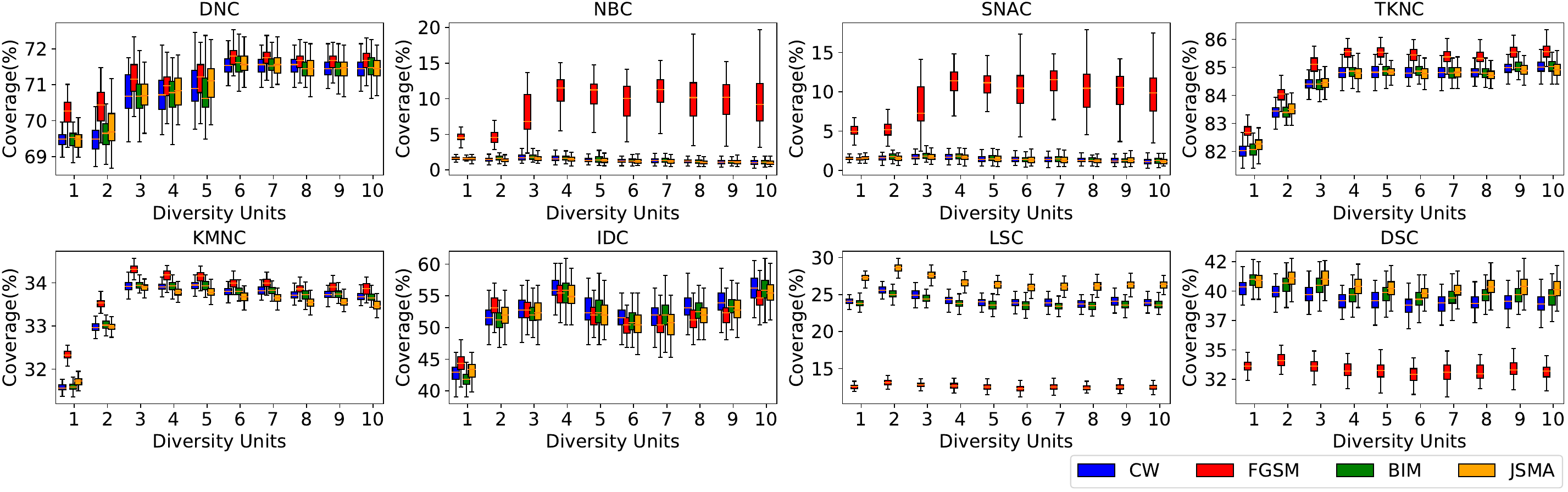}
 \end{tabular}
 \vspace{-3mm}
 \caption{Correlation between test coverage and test diversity on the subject (ID: 4).}
 \label{fig:rq2_cifar10_resnet20}
\end{figure*}

\subsection{RQ2: Correlation between Test Coverage and Test Diversity}

\begin{wraptable}[19]{l}{0.47\textwidth}
\vspace{-5mm}
    \caption{\label{tab:rq2cor} RQ2: Spearman correlation between test coverage and test diversity. }
    \begin{adjustbox}{max width=0.5\textwidth,left}
    \begin{threeparttable}
    \begin{tabular}{clrrrrrrrr}
    \headrow
    \thead{ID} & \thead{Adv.} & \thead{DNC} & \thead{NBC} & \thead{SNAC} & \thead{TKNC} & \thead{KMNC} & \thead{IDC} & \thead{LSC} & \thead{DSC} \\\hline
    \hiderowcolors
    \multirow{4}{*}{1} & CW & \cellcolor{MyGreen!100!black}\textbf{0.99} & \cellcolor{MyGreen!50!white}\textbf{0.75} & \cellcolor{MyGreen!50!white}\textbf{0.81} & \cellcolor{MyGreen!100!black}\textbf{0.99} & 0.62 & \cellcolor{MyGreen!100!black}\textbf{0.94} & 0.01 & 0.43 \\
 & FGSM & \cellcolor{MyGreen!100!black}\textbf{0.99} & \cellcolor{MyGreen!15!white}\textbf{0.66} & \cellcolor{MyGreen!50!white}\textbf{0.83} & \cellcolor{MyGreen!100!black}\textbf{0.98} & \cellcolor{MyGreen!50!white}\textbf{0.77} & \cellcolor{MyGreen!100!black}\textbf{0.99} & 0.25 & \cellcolor{MyGreen!50!white}\textbf{0.83} \\
 & BIM & \cellcolor{MyGreen!100!black}\textbf{0.99} & 0.15 & \cellcolor{MyGreen!15!white}\textbf{0.66} & \cellcolor{MyGreen!100!black}\textbf{0.98} & 0.62 & \cellcolor{MyGreen!100!black}\textbf{0.95} & 0.02 & 0.54 \\
 & JSMA & \cellcolor{MyGreen!100!black}\textbf{0.99} & -0.41 & \cellcolor{MyGreen!50!white}\textbf{0.76} & \cellcolor{MyGreen!100!black}\textbf{0.99} & 0.62 & \cellcolor{MyGreen!100!black}\textbf{0.95} & -0.12 & 0.43 \\ \hline
\multirow{4}{*}{2} & CW & \cellcolor{MyGreen!100!black}\textbf{0.93} & 0.2 & 0.14 & \cellcolor{MyGreen!100!black}\textbf{0.96} & \cellcolor{MyGreen!50!white}\textbf{0.88} & \cellcolor{MyGreen!50!white}\textbf{0.87} & -0.15 & -0.15 \\ 
 & FGSM & \cellcolor{MyGreen!100!black}\textbf{0.96} & \cellcolor{MyGreen!15!white}\textbf{0.7} & \cellcolor{MyGreen!15!white}\textbf{0.7} & \cellcolor{MyGreen!100!black}\textbf{0.96} & \cellcolor{MyGreen!50!white}\textbf{0.84} & \cellcolor{MyGreen!50!white}\textbf{0.87} & 0.2 & \cellcolor{MyGreen!50!white}\textbf{0.82} \\
 & BIM & \cellcolor{MyGreen!100!black}\textbf{0.99} & 0.41 & 0.41 & \cellcolor{MyGreen!100!black}\textbf{0.98} & \cellcolor{MyGreen!50!white}\textbf{0.85} & \cellcolor{MyGreen!50!white}\textbf{0.88} & -0.15 & -0.38 \\
 & JSMA & \cellcolor{MyGreen!100!black}\textbf{0.96} & 0.49 & 0.08 & \cellcolor{MyGreen!100!black}\textbf{0.98} & \cellcolor{MyGreen!50!white}\textbf{0.85} & \cellcolor{MyGreen!50!white}\textbf{0.88} & 0.2 & 0.42 \\ \hline
\multirow{4}{*}{3} & CW & \cellcolor{MyGreen!100!black}\textbf{0.9} & \textbf{-0.75} & \textbf{-0.75} & \cellcolor{MyGreen!100!black}\textbf{1.0} & 0.33 & \cellcolor{MyGreen!50!white}\textbf{0.83} & -0.36 & 0.14 \\ 
 & FGSM & \cellcolor{MyGreen!15!white}\textbf{0.65} & -0.03 & -0.03 & \cellcolor{MyGreen!100!black}\textbf{0.99} & 0.03 & \cellcolor{MyGreen!50!white}\textbf{0.84} & -0.27 & -0.26 \\
 & BIM & \cellcolor{MyGreen!50!white}\textbf{0.75} & \textbf{-0.82} & \textbf{-0.82} & \cellcolor{MyGreen!100!black}\textbf{1.0} & 0.22 & \cellcolor{MyGreen!50!white}\textbf{0.87} & -0.27 & -0.15 \\
 & JSMA & \cellcolor{MyGreen!50!white}\textbf{0.89} & \textbf{-0.71} & \textbf{-0.71} & \cellcolor{MyGreen!100!black}\textbf{1.0} & 0.07 & \cellcolor{MyGreen!50!white}\textbf{0.84} & -0.26 & 0.5 \\ \hline
\multirow{4}{*}{4} & CW & \cellcolor{MyGreen!50!white}\textbf{0.75} & \textbf{-0.9} & \textbf{-0.88} & \cellcolor{MyGreen!50!white}\textbf{0.85} & 0.04 & \cellcolor{MyGreen!15!white}\textbf{0.7} & \textbf{-0.75} & \textbf{-0.78} \\ 
 & FGSM & \cellcolor{MyGreen!50!white}\textbf{0.76} & 0.45 & 0.45 & \cellcolor{MyGreen!50!white}\textbf{0.72} & 0.03 & 0.12 & \textbf{-0.7} & \textbf{-0.64} \\
 & BIM & \cellcolor{MyGreen!50!white}\textbf{0.79} & \textbf{-0.87} & \textbf{-0.82} & \cellcolor{MyGreen!50!white}\textbf{0.81} & 0.02 & \cellcolor{MyGreen!15!white}\textbf{0.7} & -0.62 & \textbf{-0.72} \\
 & JSMA & \cellcolor{MyGreen!50!white}\textbf{0.79} & \textbf{-0.93} & \textbf{-0.88} & \cellcolor{MyGreen!50!white}\textbf{0.82} & 0.01 & 0.45 & -0.61 & \textbf{-0.67} \\ \hline
\multirow{4}{*}{5} & CW & \cellcolor{MyGreen!100!black}\textbf{0.98} & \textbf{-0.93} & \textbf{-0.93} & \cellcolor{MyGreen!50!white}\textbf{0.89} & 0.36 & 0.15 & -0.61 & -0.52 \\ 
 & FGSM & \cellcolor{MyGreen!100!black}\textbf{0.98} & -0.13 & -0.13 & \cellcolor{MyGreen!50!white}\textbf{0.85} & 0.35 & 0.05 & -0.46 & 0.32 \\
 & BIM & \cellcolor{MyGreen!100!black}\textbf{0.98} & -0.43 & -0.43 & \cellcolor{MyGreen!50!white}\textbf{0.85} & \cellcolor{MyGreen!15!white}\textbf{0.66} & 0.13 & -0.59 & \cellcolor{MyGreen!100!black}\textbf{0.92} \\
 & JSMA & \cellcolor{MyGreen!100!black}\textbf{0.98} & \textbf{-0.94} & \textbf{-0.95} & \cellcolor{MyGreen!100!black}\textbf{0.92} & 0.25 & 0.14 & \textbf{-0.83} & -0.2 \\ \hline
\multirow{4}{*}{6} & CW & \cellcolor{MyGreen!100!black}\textbf{0.98} & 0.53 & 0.56 & \cellcolor{MyGreen!100!black}\textbf{0.95} & \cellcolor{MyGreen!100!black}\textbf{0.98} & 0.6 & -0.18 & -0.37 \\ 
 & FGSM & \cellcolor{MyGreen!100!black}\textbf{0.99} & 0.43 & 0.45 & \cellcolor{MyGreen!100!black}\textbf{0.96} & \cellcolor{MyGreen!100!black}\textbf{0.99} & 0.6 & 0.05 & \textbf{-0.68} \\
 & BIM & \cellcolor{MyGreen!100!black}\textbf{0.98} & \cellcolor{MyGreen!50!white}\textbf{0.81} & \cellcolor{MyGreen!50!white}\textbf{0.82} & \cellcolor{MyGreen!100!black}\textbf{0.98} & \cellcolor{MyGreen!100!black}\textbf{0.98} & \cellcolor{MyGreen!15!white}\textbf{0.65} & -0.2 & -0.29 \\
 & JSMA & \cellcolor{MyGreen!100!black}\textbf{1.0} & 0.53 & 0.53 & \cellcolor{MyGreen!100!black}\textbf{0.98} & \cellcolor{MyGreen!100!black}\textbf{0.96} & \cellcolor{MyGreen!50!white}\textbf{0.71} & 0.13 & -0.26 \\
    
    \hline  
    \end{tabular}
    \end{threeparttable}
    \end{adjustbox}
    \vspace{5mm}
\end{wraptable}
For each coverage on each subject, we plot the correlation between test coverage and test diversity, where the x-axis represents the ratio of covered classes (representing test diversity) and the y-axis represents the achieved coverage.
Different colors represent the results of different adversarial test input generation methods.
We found that the conclusions across different subjects are similar and due to space limit, we show the results of CIFAR-10 and ResNet-20 (ID: 4) as the representative in Figure~\ref{fig:rq2_cifar10_resnet20}.
All the results can be found on our project homepage~\cite{homepage}.

In particular, we calculated the Spearman correlation between test coverage and test diversity by using the medium of each box as the achieved coverage under the corresponding degree of test diversity for calculation, as shown in Table~\ref{tab:rq2cor}.
According to these tables and figures, we obtained the following major conclusions.

First, regarding the four structural coverage metrics (i.e., DNC, TKNC, KMNC, and IDC), with the test diversity increasing, the achieved coverage increases (especially when the test diversity is small).
That indicates when controlling for the size and the ratio of error-revealing test inputs, they show positive correlation (especially for DNC and TKNC).
That is, different neurons are responsible for the prediction of different classes.
When covering more classes, more neurons could be activated, leading to the increment of these structural coverage metrics.

When the test diversity reaches a certain extent, the increment is starting to flatten.
The possible reason lies in the limitation of the test-set size.
When the number of covered classes becomes large in a test set, the number of test inputs for each class becomes small accordingly, which may cause the responsible neurons for each class are not sufficiently activated. 
That is, the coverage increment brought by increasing test diversity may be counteracted by the coverage decrement caused by decreasing the number of test inputs for each class. 


Second, regarding the remaining two structural coverage metrics (i.e., NBC and SNAC), in most cases there is no obvious positive correlation between the two metrics and test diversity.
The correlation tends to be stronger when using FGSM to generate adversarial test inputs.
The reason is that it is difficult to improve both NBC and SNAC due to their inherent ``corner-case'' characteristics, and thus regardless of test diversity, both NBC and SNAC stay small values stably.
Regarding the results on FGSM, as presented before, the adversarial test inputs generated by FGSM are more likely to cover corner-case regions of the output values for neurons, causing that both NBC and SNAC may break away from small values.

Third, regarding the two non-structural coverage metrics, the achieved coverage does not increase with the test diversity increasing, indicating no obvious positive correlation between them.
This is because these metrics measure the distance between a test input and the training data with the same label.
Since training data tend to be sufficient, the distributions of the training data for different classes are stable, causing that the distance cannot be affected obviously by different classes of test inputs.

\subsection{RQ3: Correlation between Test Coverage and Error-revealing Capability per Diversity Unit}

\begin{wraptable}[18]{l}{0.47\textwidth}
\centering
\vspace{-5mm}
\caption{\label{tab:rq3_10percent} RQ3: Spearman correlation between test coverage and error-revealing capability per diversity unit (1$^{st}$ group).}
\begin{adjustbox}{width=0.5\textwidth,left}
\begin{threeparttable}
\begin{tabular}{clrrrrrrrr}
\headrow
\thead{ID} & \thead{Adv.} & \thead{DNC} & \thead{NBC} & \thead{SNAC} & \thead{TKNC} & \thead{KMNC} & \thead{IDC} & \thead{LSC} & \thead{DSC} \\
\hline
\hiderowcolors
\multirow{4}{*}{1} & CW & \cellcolor{MyGreen!50!white}\textbf{0.72} & \textbf{-0.43} & \textbf{-0.43} & \cellcolor{MyGreen!50!white}\textbf{0.8} & \cellcolor{MyGreen!50!white}\textbf{0.79} & \cellcolor{MyGreen!50!white}\textbf{0.77} & \cellcolor{MyGreen!100!black}\textbf{0.98} & \cellcolor{MyGreen!100!black}\textbf{0.99} \\
 & FGSM & \cellcolor{MyGreen!50!white}\textbf{0.76} & \cellcolor{MyGreen!15!white}\textbf{0.46} & \textbf{0.32} & \cellcolor{MyGreen!50!white}\textbf{0.86} & \cellcolor{MyGreen!15!white}\textbf{0.53} & 0.06 & \cellcolor{MyGreen!100!black}\textbf{0.97} & \cellcolor{MyGreen!100!black}\textbf{0.97} \\
 & BIM & \cellcolor{MyGreen!50!white}\textbf{0.8} & \textbf{-0.47} & \textbf{-0.47} & \cellcolor{MyGreen!50!white}\textbf{0.9} & \cellcolor{MyGreen!50!white}\textbf{0.72} & \cellcolor{MyGreen!50!white}\textbf{0.77} & \cellcolor{MyGreen!100!black}\textbf{0.97} & \cellcolor{MyGreen!100!black}\textbf{0.97} \\
 & JSMA & \cellcolor{MyGreen!50!white}\textbf{0.85} & \textbf{-0.46} & \textbf{-0.46} & \cellcolor{MyGreen!50!white}\textbf{0.89} & \cellcolor{MyGreen!100!black}\textbf{0.96} & \cellcolor{MyGreen!15!white}\textbf{0.68} & \cellcolor{MyGreen!100!black}\textbf{0.99} & \cellcolor{MyGreen!100!black}\textbf{0.99} \\
 \hline
\multirow{4}{*}{2} & CW & \cellcolor{MyGreen!50!white}\textbf{0.86} & \textbf{0.39} & 0.04 & \cellcolor{MyGreen!50!white}\textbf{0.79} & \cellcolor{MyGreen!50!white}\textbf{0.71} & \cellcolor{MyGreen!15!white}\textbf{0.44} & \cellcolor{MyGreen!50!white}\textbf{0.87} & \cellcolor{MyGreen!100!black}\textbf{0.94} \\
 & FGSM & \cellcolor{MyGreen!50!white}\textbf{0.8} & \cellcolor{MyGreen!15!white}\textbf{0.43} & \textbf{0.23} & \cellcolor{MyGreen!50!white}\textbf{0.84} & \cellcolor{MyGreen!50!white}\textbf{0.75} & \cellcolor{MyGreen!15!white}\textbf{0.42} & \cellcolor{MyGreen!50!white}\textbf{0.79} & \cellcolor{MyGreen!50!white}\textbf{0.88} \\
 & BIM & \cellcolor{MyGreen!50!white}\textbf{0.77} & \textbf{0.23} & 0.04 & \cellcolor{MyGreen!50!white}\textbf{0.73} & \cellcolor{MyGreen!15!white}\textbf{0.43} & \textbf{0.36} & \cellcolor{MyGreen!50!white}\textbf{0.81} & \cellcolor{MyGreen!100!black}\textbf{0.94} \\
 & JSMA & \cellcolor{MyGreen!50!white}\textbf{0.84} & \cellcolor{MyGreen!15!white}\textbf{0.54} & \textbf{0.32} & \cellcolor{MyGreen!50!white}\textbf{0.8} & \cellcolor{MyGreen!50!white}\textbf{0.8} & \cellcolor{MyGreen!15!white}\textbf{0.5} & \cellcolor{MyGreen!50!white}\textbf{0.82} & \cellcolor{MyGreen!100!black}\textbf{0.95} \\
 \hline
\multirow{4}{*}{3} & CW & \cellcolor{MyGreen!100!black}\textbf{0.94} & \textbf{-0.6} & \textbf{-0.65} & \cellcolor{MyGreen!100!black}\textbf{0.98} & \cellcolor{MyGreen!100!black}\textbf{0.96} & \cellcolor{MyGreen!15!white}\textbf{0.61} & \cellcolor{MyGreen!100!black}\textbf{0.95} & \cellcolor{MyGreen!15!white}\textbf{0.5} \\
 & FGSM & \cellcolor{MyGreen!100!black}\textbf{0.95} & \cellcolor{MyGreen!50!white}\textbf{0.87} & \cellcolor{MyGreen!50!white}\textbf{0.86} & \cellcolor{MyGreen!100!black}\textbf{0.98} & \cellcolor{MyGreen!100!black}\textbf{0.99} & \cellcolor{MyGreen!15!white}\textbf{0.67} & \cellcolor{MyGreen!50!white}\textbf{0.88} & \cellcolor{MyGreen!100!black}\textbf{0.94} \\
 & BIM & \cellcolor{MyGreen!100!black}\textbf{0.95} & \cellcolor{MyGreen!50!white}\textbf{0.8} & \cellcolor{MyGreen!50!white}\textbf{0.8} & \cellcolor{MyGreen!100!black}\textbf{0.98} & \cellcolor{MyGreen!100!black}\textbf{0.97} & \cellcolor{MyGreen!50!white}\textbf{0.73} & \cellcolor{MyGreen!100!black}\textbf{0.94} & \cellcolor{MyGreen!50!white}\textbf{0.89} \\
 & JSMA & \cellcolor{MyGreen!100!black}\textbf{0.96} & \textbf{-0.57} & \textbf{-0.69} & \cellcolor{MyGreen!100!black}\textbf{0.99} & \cellcolor{MyGreen!100!black}\textbf{0.96} & \cellcolor{MyGreen!15!white}\textbf{0.47} & \cellcolor{MyGreen!100!black}\textbf{0.92} & \cellcolor{MyGreen!15!white}\textbf{0.67} \\
 \hline
\multirow{4}{*}{4} & CW & \cellcolor{MyGreen!15!white}\textbf{0.63} & -0.19 & -0.09 & \cellcolor{MyGreen!50!white}\textbf{0.85} & \textbf{0.32} & \cellcolor{MyGreen!100!black}\textbf{0.96} & \cellcolor{MyGreen!50!white}\textbf{0.84} & \cellcolor{MyGreen!50!white}\textbf{0.73} \\
 & FGSM & \cellcolor{MyGreen!15!white}\textbf{0.63} & \cellcolor{MyGreen!50!white}\textbf{0.8} & \cellcolor{MyGreen!50!white}\textbf{0.85} & \cellcolor{MyGreen!50!white}\textbf{0.89} & \cellcolor{MyGreen!100!black}\textbf{0.93} & \cellcolor{MyGreen!100!black}\textbf{0.94} & \cellcolor{MyGreen!50!white}\textbf{0.83} & \cellcolor{MyGreen!15!white}\textbf{0.62} \\
 & BIM & \textbf{0.38} & -0.06 & 0.13 & \cellcolor{MyGreen!50!white}\textbf{0.85} & \cellcolor{MyGreen!15!white}\textbf{0.6} & \cellcolor{MyGreen!100!black}\textbf{0.94} & \cellcolor{MyGreen!50!white}\textbf{0.82} & \cellcolor{MyGreen!15!white}\textbf{0.62} \\
 & JSMA & \cellcolor{MyGreen!15!white}\textbf{0.41} & \textbf{-0.22} & -0.09 & \cellcolor{MyGreen!50!white}\textbf{0.84} & \cellcolor{MyGreen!15!white}\textbf{0.45} & \cellcolor{MyGreen!100!black}\textbf{0.94} & \cellcolor{MyGreen!50!white}\textbf{0.86} & \cellcolor{MyGreen!15!white}\textbf{0.51} \\
 \hline
\multirow{4}{*}{5} & CW & \cellcolor{MyGreen!50!white}\textbf{0.84} & \textbf{-0.56} & \textbf{-0.54} & \cellcolor{MyGreen!50!white}\textbf{0.89} & \textbf{-0.72} & \textbf{-0.33} & \cellcolor{MyGreen!15!white}\textbf{0.55} & \cellcolor{MyGreen!100!black}\textbf{0.95} \\
 & FGSM & \cellcolor{MyGreen!50!white}\textbf{0.73} & \cellcolor{MyGreen!15!white}\textbf{0.5} & \cellcolor{MyGreen!15!white}\textbf{0.49} & \cellcolor{MyGreen!100!black}\textbf{0.91} & \textbf{-0.53} & \textbf{-0.48} & -0.04 & \cellcolor{MyGreen!100!black}\textbf{0.95} \\
 & BIM & \cellcolor{MyGreen!15!white}\textbf{0.68} & \cellcolor{MyGreen!15!white}\textbf{0.63} & \cellcolor{MyGreen!15!white}\textbf{0.64} & \cellcolor{MyGreen!50!white}\textbf{0.89} & \textbf{-0.85} & \textbf{-0.48} & 0.04 & \cellcolor{MyGreen!100!black}\textbf{0.95} \\
 & JSMA & \cellcolor{MyGreen!50!white}\textbf{0.72} & \textbf{-0.5} & \textbf{-0.5} & \cellcolor{MyGreen!100!black}\textbf{0.92} & \textbf{-0.77} & \textbf{-0.37} & \textbf{0.22} & \cellcolor{MyGreen!100!black}\textbf{0.95} \\
 \hline
\multirow{4}{*}{6} & CW & \cellcolor{MyGreen!15!white}\textbf{0.67} & -0.08 & -0.05 & 0.12 & \textbf{-0.53} & \cellcolor{MyGreen!15!white}\textbf{0.5} & 0.19 & \textbf{-0.85} \\
 & FGSM & \cellcolor{MyGreen!50!white}\textbf{0.76} & \cellcolor{MyGreen!15!white}\textbf{0.64} & \cellcolor{MyGreen!15!white}\textbf{0.63} & \cellcolor{MyGreen!15!white}\textbf{0.68} & \cellcolor{MyGreen!15!white}\textbf{0.66} & \cellcolor{MyGreen!15!white}\textbf{0.65} & \cellcolor{MyGreen!50!white}\textbf{0.83} & \textbf{-0.63} \\
 & BIM & \cellcolor{MyGreen!15!white}\textbf{0.69} & \textbf{-0.24} & \textbf{-0.21} & 0.15 & \textbf{-0.42} & \cellcolor{MyGreen!15!white}\textbf{0.67} & \textbf{0.26} & \textbf{-0.65} \\
 & JSMA & \cellcolor{MyGreen!50!white}\textbf{0.71} & 0.14 & 0.14 & 0.01 & \textbf{-0.37} & \cellcolor{MyGreen!15!white}\textbf{0.6} & \cellcolor{MyGreen!15!white}\textbf{0.57} & \textbf{-0.78} \\ 

 \hline
\end{tabular}
\end{threeparttable}
\end{adjustbox}
\end{wraptable}
According to RQ1 and RQ2, we found that some of test coverage metrics are positively correlated with the test diversity, but are not positively correlated with error-revealing capability under saturated diversity. 
One subsequent question is that when test diversity is not saturated, what about the correlation between test coverage and error-revealing capability?

\label{sec:structure}
\begin{figure*}[t!]
 \center
 \hspace{-2mm}
 \begin{tabular}{l}
  \includegraphics[scale=0.3]{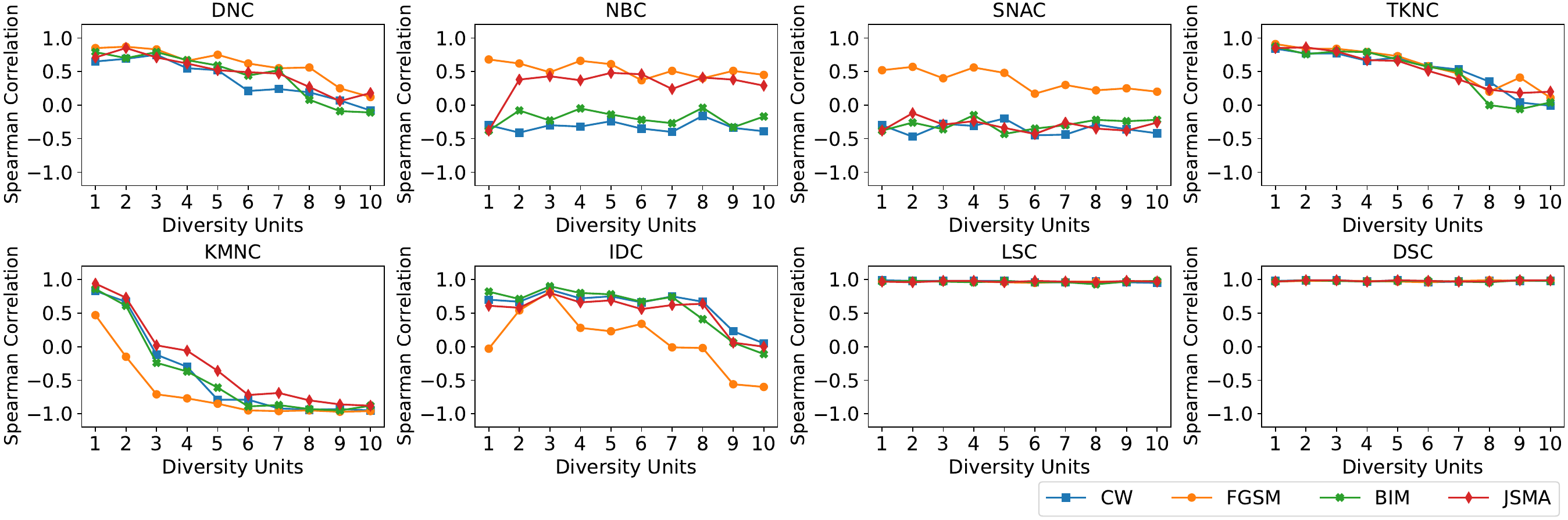}
 \end{tabular}
 \vspace{-5mm}
 \caption{Trend of correlation between coverage and error-revealing capability with test diversity increasing on subject (ID: 1)}
 \label{fig:rq3}
 \vspace{-5mm}
\end{figure*}

To answer this question, for each coverage on each subject, we measured the correlation between test coverage and error-revealing capability per diversity unit.
Although we repeated the experiment five times by selecting different groups to construct the test sets with one diversity unit (presented in Section~\ref{rq3}), we only show the results on the first group in Table~\ref{tab:rq3_10percent} due to the space limit.

Indeed, the conclusions are consistent across different groups, which can be found on our project homepage~\cite{homepage}.
From Table~\ref{tab:rq3_10percent}, we found that the four structural coverage metrics (DNC, TKNC, KMNC, and IDC) that are positively correlated with test diversity, also have positive correlation with error-revealing capability under small test diversity (i.e., one diversity unit).
However, as shown in RQ1, they have no correlation or negative correlation with error-revealing capability under saturated diversity. 
The reason is that although the test inputs in a test set concentrate on only 10\% of classes, the error-revealing ones among them may predict the classes out of the 10\% of classes, causing that they are more likely to share similar traces in the DNN with the other classes of test inputs.
That is, under such a small degree of test diversity, increasing the ratio of error-revealing test inputs may cover different neurons with those covered by the 10\% of classes of test inputs.



Furthermore, we investigated the correlation trend with the degree of test diversity increasing from one diversity unit to saturated diversity (i.e., 10 diversity units) with the step size of a diversity unit.
Due to the space limit, we show the trend results by taking MNIST and LeNet-5 (ID: 1) as the representative in Figure~\ref{fig:rq3}.
Indeed, there are almost consistent conclusions across different subjects, and we put all the results on our project homepage~\cite{homepage}.
In Figure~\ref{fig:rq3}, the x-axis represents the degree of test diversity and the y-axis represents the Spearman correlation coefficient. 
From Figure~\ref{fig:rq3}, with the degree of test diversity increasing, the correlation between test coverage and error-revealing capability becomes weaker for the four structural coverage metrics (DNC, TKNC, KMNC, and IDC).
This is as expected, since when the degree of test diversity is large, more error-revealing test inputs cannot bring more coverage of different neurons that are covered by other classes.
Therefore, the four structural coverage metrics are not useless to guide the generation of test inputs, but the proper usage scenario for them is not found before.
In actual, they can be effective to guide to generate error-revealing test inputs when using a set of test inputs with small test diversity as seed inputs.

Regarding the remaining four test coverage metrics (i.e., NBC, SNAC, LSC, and DSC), with the degree of test diversity increasing, the correlation between test coverage and error-revealing capability is stable.
That is, their correlation between test coverage and error-revealing capability is not largely affected by various degrees of test diversity, which is as expected based on the conclusions obtained in RQ1 and RQ2. 
Specifically, the correlation is stably strong for LSC and DSC, but is stably weak for NBC and SNAC.

%

\vspace{5mm}
\subsection{Conclusion}

In this paper, we revisited the existing DNN test coverage from the test-effectiveness perspective by considering three test-effectiveness criteria (i.e., error-revealing capability, test diversity, and error-revealing capability per diversity unit). Different from those previous studies, we controlled the size of test sets to more reasonably validate the correlation between DNN test coverage and the three test-effectiveness criteria. In the following, we summarize the main findings by addressing three research questions in our study:

\noindent\textbf{Findings for RQ1}: (1) In general, the six structural coverage metrics do not have moderate to extremely strong positive correlation with the error-revealing capability of the test set while the two non-structural coverage metrics have, when controlling for the size and saturated diversity of test sets, showing the superiority of non-structural coverage in this scenario. (2) Though non-structural coverage is more effective in perceiving the error-revealing test inputs, the parameters of LSC and DSC significantly impact their performance. How to set the parameters properly remains a challenge. (3) Compared with the classification models, the correlation on the regression model is stronger since the metric used to measure the accuracy of regression (i.e., MSE) is more fine-grained and sensitive to capturing neuron coverage differences. 

\noindent\textbf{Findings for RQ2}: (1) When controlling for the size and the ratio of the error-revealing test inputs, four structural coverage metrics (i.e., DNC, TKNC, KMNC, and IDC) have positive correlation with test diversity, especially under small test diversity, demonstrating that the four structural coverage metrics can guide the improvement of test diversity. (2) The remaining two structural coverage metrics (i.e., NBC and SNAC) and the non-structural coverage metrics are not significantly correlated to test diversity. 

\noindent\textbf{Findings for RQ3}: (1) The four structural coverage metrics (i.e., DNC, TKNC, KMNC, and IDC) are positively correlated with error-revealing capability under small test diversity, and the correlation becomes weaker with the degree of test diversity increasing, indicating that they are useful to guide the generation of error-revealing test inputs by taking the test inputs with small test diversity as seed inputs. (2) As for the remaining four test coverage metrics (i.e., NBC, SNAC, LSC, and DSC), their correlation between test coverage and error-revealing capability is relatively stable and not largely affected by various degrees of test diversity.

\vspace{-1.5mm}
\section{Implications and Suggestions}
\label{sec:implications}
\vspace{-3mm}


Based on the conclusions from our extensive study, we summarize a series of implications from both academia and practice perspectives. 

\vspace{0.5mm}
\noindent\textbf{Implications from the academia perspective}:
(1) Some empirical studies have been conducted to investigate the usefulness of existing DNN test coverage from some perspectives, and they delivered negative conclusions that may drive future research towards the direction away from the existing coverage.
Different from them, our work investigates the usefulness of these coverage metrics from another perspective (i.e., test effectiveness measured by different criteria), which identifies the practical usage scenarios for the existing coverage.
That is, our study can inspire future research that continue around the \textit{existing} coverage but in their proper scenarios.
In recent years, many studies have been conducted to overturn a (small) area of existing techniques/measurements, opening another direction for the area.
They are appealing, but revisiting these existing techniques/measurements from different perspectives and identifying the proper usage scenarios for them are also inspiring for the development of the area. Please note that we do not overturn the conclusions from the previous studies since those previous studies and our study were conducted from different perspectives and used different criteria to measure the correlation with DNN test coverage. Our test-effectiveness perspective and criteria help identify the usage scenarios of the existing DNN test coverage, making the overall conclusions positive to a large extent.

(2) The usefulness of some structural coverage metrics can be manifested by incorporating test diversity.
In our study, we propose a coarse-grained measurement (i.e., the ratio of covered classes among all classes involved in a DNN) for test diversity, but such a coarse-grained measurement can indeed inspire the usefulness of these coverage metrics.
In the future, it is promising to design more fine-grained measurements for test diversity in order to better utilize the existing structural coverage (such as measuring test-input diversity under each class based on input features).
Moreover, it is also important to incorporate the concept of test diversity in regression tasks.

(3) The correlation between test coverage and error-revealing capability is much stronger on regression models than classification models due to the fine-grained accuracy measurement for the former.
Such a fine-grained measurement is more sensitive to capture the differences in test behaviors (such as neuron coverage differences).
Therefore, decomposing the accuracy measurement for classification models to be more fine-grained may be a promising direction to improve the usefulness of these coverage metrics (especially structural coverage).

\begin{wrapfigure}[11]{l}{0.42\textwidth}
 \centering
 \vspace{-3mm}
  \includegraphics[scale=0.32]{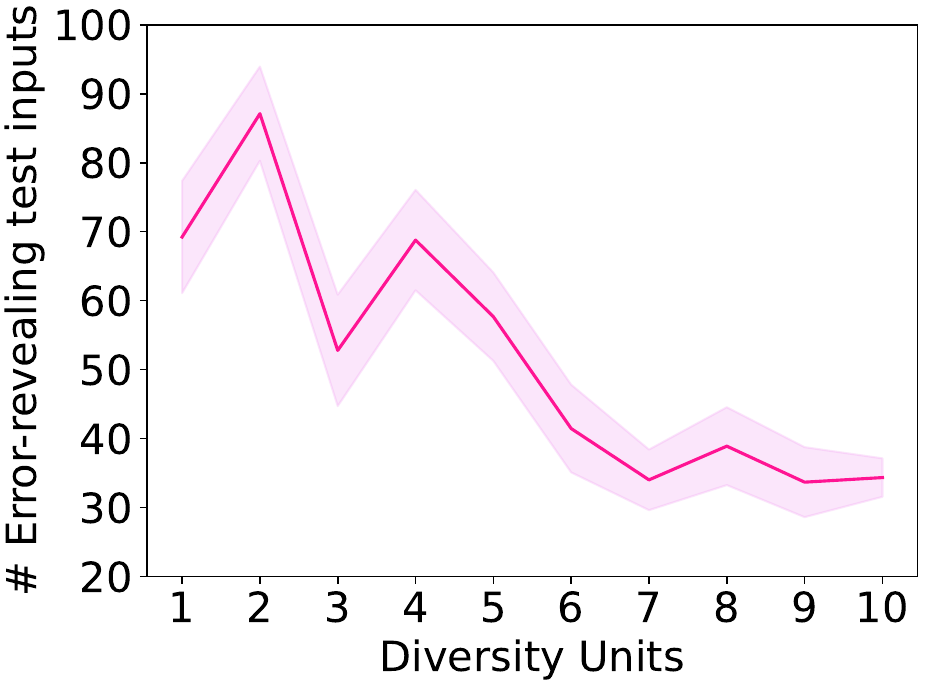}
\vspace{-3mm}
 \caption{\label{fig:cov_attack} Trend of attack success rate with test diversity increasing on the subject (ID: 1).}
\end{wrapfigure}
\noindent\textbf{Implications from the practice perspective}:
(1) The usage scenario of the non-structural coverage (i.e., LSC and DSC) is using it to guide the generation of error-revealing test inputs regardless of taking the whole test set or a small number of classes of test inputs as seed inputs.
But the remarkable thing is how to properly set the parameters of LSC and DSC for different subjects, which can be studied carefully in the future.
Moreover, it is hard to guarantee the diversity of the generated test inputs guided by LSC and DSC.

(2) The usage scenario of the structural coverage (i.e., DNC, TKNC, KMNC, and IDC) is using it to guide the generation of error-revealing test inputs by taking a small number of classes of test inputs as seed inputs.
In particular, the guidance of these structural coverage can also help improve the diversity of the generated test inputs.
However, they are not capable of guiding to generate error-revealing test inputs by taking the whole test set (with saturated diversity) as seed inputs. 
We conducted a proof-of-concept application of our finding to further make our implication actionable. Here, we applied DNC to guide the generation of test inputs on MNIST-LeNet5 under different degrees of test diversity. Specifically, for each degree of test diversity, we randomly sampled 200 test inputs as seeds and then used DNC to guide the generation of test inputs by maximizing activation values of inactive neurons and taking the penultimate fully-connected layer as the targeted layer. The maximum iteration is set to 50 and the process was repeated 10 times to reduce the influence of randomness. The Figure~\ref{fig:cov_attack} shows the results, where the x-axis represents different degrees of test diversity while the y-axis represents the number of successfully generated error-revealing test inputs. We find that DNC can help generate more error-revealing test inputs when seeds have small test diversity. This is a specific scenario that is suiteable for structural coverage.



(3) In practice, we can first use the four structural coverage metrics (i.e., DNC, TKNC, KMNC, and IDC) to guide the generation of \textit{diverse} error-revealing test inputs by taking each class of test inputs as seed inputs respectively.
When the generated error-revealing test inputs based on each class of test inputs are diverse enough, we can then adopt the non-structural coverage (i.e., LSC and DSC) to guide to generate \textit{more} error-revealing test inputs by taking diverse test inputs as seed inputs, which can complement with the four structural coverage metrics.




\vspace{-4mm}
\section{Discussion}
\label{sec:discuss} 
\vspace{-4mm}

Apart from the six structural coverage studied above, Sun et al.~\cite{DBLP:journals/tecs/SunHKSHA19} proposed MC/DC-inspired neuron coverage, which is tailored to the distinct features of a DNN.
It instantiates the decision change and condition change in a DNN by defining the \textit{sign change} and \textit{value change} of neurons, which show the degree of the change of neuron activation values under two different test inputs.
Same as the existing studies on DNN coverage~\cite{harel2020neuron,yan2020correlations}, we did not investigate MC/DC-inspired neuron coverage~\cite{DBLP:conf/icse/SunHKSHA19} in our study, since it is limited in scalability (hard to apply to large subjects) as demonstrated by the existing work~\cite{ma2018deepgauge,zhou2021deepcon}.

Indeed, it is interesting to investigate whether our conclusions still hold for this state-of-the-art structural coverage. Thus, we conducted an experiment by taking VGG-16, CIFAR-10, C\&W, and the MC/DC-inspired coverage metric – Sign-Sign Coverage, as the representative. We repeated the process in RQ1 by constructing 200 test sets with the size of 1000. By calculating the Spearman correlation between the coverage and the ratio of error-revealing test inputs, we found that the correlation coefficient is -0.54 and the p-value is 5.92e-18 (< 0.05). The result demonstrates there is negative correlation between the Sign-Sign Coverage and the ratio of error-revealing test inputs, which is consistent with the conclusions from the other studied structural coverage, indicating the generality of our conclusions to some degree.

\vspace{-1mm}
\section{Threats to Validity}
\label{sec:threat} 
\vspace{-4mm}


Our study mainly has five kinds of threats to \textit{external} validity.
(1) \textit{Subjects}: Our used subjects may not represent other subjects. To reduce this kind of threat, we carefully considered the diversity of subjects and used some new subjects that have never been used to evaluate these studied DNN test coverage metrics.
(2) \textit{Created test sets}: We controlled for the size of created test sets, i.e., 800, which may not represent other sizes.
Actually, we have conducted a small experiment to investigate the influence of different test-set sizes (i.e., 100, 500, 800, and 1000) and obtained almost consistent conclusions.
We put the experimental results on our project homepage~\cite{homepage}.
(3) \textit{Studied test coverage}: Our studied test coverage may not represent other coverage. To reduce this kind of threat, we considered all the DNN test coverage studied in the existing studies~\cite{yan2020correlations,harel2020neuron} (except TKNP~\cite{ma2018deepgauge}) and additionally studied the most recent test coverage.
In the future, we will study more coverage metrics, such as t-way combination coverage~\cite{ma2019deepct} and state coverage~\cite{du2018deepcruiser}.
(4) \textit{Adversarial test input generation methods}: We adopted the most widely-used FGSM, BIM, JSMA, C\&W for general image-domain DNN models, the widely-used DeepXplore-Patch and DeepXplore-Light for Driving following the existing work~\cite{Li:2019:BOD:3338906.3338930,kim2019guiding}, CTCM for Speech-Commands and PWWS for 20-Newsgroups, which may not represent other methods.
However, this kind of threat may be not serious since the conclusions obtained from BIM, JSMA, and C\&W are almost consistent.
(5)\textit{Used errors}: We regard a test input predicted wrongly as an error-revealing test input, but do not distinguish whether they trigger the same bugs.
This kind of threat exists in all existing studies to evaluate DNN coverage since it is challenging to distinguish them due to involving huge manual effort.
Actually, the number of error-revealing test inputs is also important in DNN testing since identifying them can help improve the accuracy of the DNN model~\cite{DBLP:conf/sigsoft/MaLLZG18}.








Our study also has a kind of threat to \textit{construct} validity:
\textit{diversity unit partition method}.
In RQ2 and RQ3, we treat covering 10\% of classes in a test set as a diversity unit.
For a subject, we evenly divided all classes into 10 groups according to the \textit{default} order of the class IDs.
Also, we constructed test sets with different numbers of diversity units according to the \textit{default} order of the group IDs.
However, the method of diversity unit partition may be not a serious kind of threat.
This is because we investigated the influence of different groups when studying the correlation between test coverage and error-revealing capability per diversity unit in RQ3, and found that the conclusions are consistent.

Besides, the hardware environment can also be a potential factor affecting our study. Regarding the influence of hardware, it may be a threat to all the studies on deep learning. Same as the existing studies~\cite{yan2020correlations,harel2020neuron}, we did not systematically investigate the influence of this factor, but reported the details of our experimental environment (in Section~\ref{sec:setup}), released our artifact for replication~\cite{homepage}, and repeated our experiments five times to reduce the influence of randomness. In the future, we will repeat our study on different hardware to further reduce this threat.



\vspace{-4mm}
\section{Related Work}
\label{sec:related}
\vspace{-4mm}
\noindent\textbf{DNN Testing}.
There is a lot of research on DNN testing~\cite{sun2018testing,zhang2019machine,DBLP:journals/tecs/SunHKSHA19,DBLP:journals/corr/abs-1910-06296,DBLP:conf/icse/TianPJR18,DBLP:conf/icse/SunHKSHA19,gerasimou2020importance,DBLP:conf/issta/LeeCLO20}.
We classify them into: measuring test effectiveness, generating test inputs, and improving test efficiency.

Regarding measuring test effectiveness, we have introduced six structural coverage metrics and two non-structural coverage metrics in Section~\ref{sec:background}, and conducted a pilot study on MC/DC inspired neuron coverage.
Besides, Ma et al.~\cite{ma2019deepct} proposed t-way combination coverage for DNN testing by borrowing the idea of traditional combinatorial testing.
Du et al.~\cite{du2018deepcruiser} proposed state coverage for RNN-based stateful DNN testing.
Ma et al.~\cite{ma2018deepmutation} proposed a mutation-based metric by designing a set of mutation operators for DNN source programs and models.
Besides the two empirical studies on DNN test coverage discussed in Section~\ref{sec:intro}, there are some other studies~\cite{li2019structural,dong2020empirical,yang2022revisiting} on evaluating DNN test coverage. 
Dong et al.~\cite{dong2020empirical} conducted a study to investigate the correlation between test coverage and model robustness and found the correlation is very limited.
Li et al.~\cite{li2019structural} conducted a preliminary study on structural coverage using only two datasets, and found that adversary-oriented search may make more contributions to the error-revealing capability than structural coverage.
Yang et al.~\cite{yang2022revisiting} extended the existing study~\cite{yan2020correlations}, further confirming the negative conclusions on the existing test coverage from the perspective of model retraining quality. 
They also claimed that the defense strategies against coverage-driven methods are worth further investigating.
Different from them, our study is to systematically revisit DNN test coverage from the test effectiveness perspective.
In particular, we designed our study systematically, i.e., considering both structural and non-structural coverage (including the state-of-the-art one), three test effectiveness criteria, and more comprehensive subjects.
Also, we did not deliver only negative conclusions like existing studies, but identified the practical usage scenarios for the existing coverage.
Furthermore, there are some metrics to measure DNN robustness~\cite{bastani2016measuring,jha2019attribution,gopinath2018deepsafe,katz2017towards}.
Since they are different from the metrics measuring test effectiveness (the target of our work), we do not discuss them.

Regarding generating test inputs, many approaches have been proposed. 
For example, Guo et al.~\cite{guo2018dlfuzz} proposed \textit{DLFuzz}, the first differential testing framework for DNNs by maximizing neuron coverage.
Xie et al.~\cite{xie2019deephunter} proposed \textit{Deephunter}, a coverage-guided fuzz testing framework for DNNs. 
Sun et al.~\cite{sun2018concolic} proposed to generate test inputs for DNNs through concolic testing. 
Shen et al.~\cite{shen2022QAQA} leveraged sentence-level mutation to generate natural test inputs to achieve precise testing of question answering software. You et al.~\cite{you2023DRFuzz} proposed DRFuzz, which aims at generating test inputs triggering regression faults for DNN models. Moreover, there are some work focusing on generating test inputs for DL libraries and DL programs~\cite{wang2020lemon,zhang2020DEBAR,yan2021grist}.
For example, Wang et al.~\cite{wang2020lemon} proposed a mutation-based approach to generating models for DL libraries. Yan et al.~\cite{yan2021grist} proposed a gradient-search-based approach to generating test inputs to expose numerical bugs in DL programs.

Regarding improving test efficiency, some test input selection and prioritization approaches have been proposed for DNNs~\cite{DBLP:journals/corr/abs-1904-13195,Li:2019:BOD:3338906.3338930,shi2019deepgini,chenpracticalTOSEM}.
For example, Li et al.~\cite{Li:2019:BOD:3338906.3338930} proposed to select test inputs by minimizing the cross entropy between the selected test inputs and the whole test set, so as to save labeling costs.
Feng et al.~\cite{shi2019deepgini} proposed \textit{DeepGini} to prioritize test inputs by measuring the purity of test inputs, so as to label error-revealing test inputs earlier.
Different from them, our work aims to investigate the correlation between DNN coverage and test effectiveness through \textit{a systematic study}.

\vspace{0.5mm}
\noindent\textbf{Empirical Studies on Traditional Test Coverage}.
In traditional software testing, there are some studies to investigate the correlation between traditional test coverage and test effectiveness.
For example, Namin and Andrews~\cite{namin2009influence} conducted a study to explore the relationship among the test-suite size, structural coverage, and test effectiveness based on C/C++ programs, and found that coverage is sometimes correlated with test effectiveness when controlling for the test-suite size.
Inozemtseva and Holmes~\cite{inozemtseva2014coverage} conducted a study to investigate the above relationship 
based on Java programs, and found that the correlation between test coverage and test effectiveness is low to moderate when controlling for the test-suite size.
Zhang and Mesbah~\cite{zhang2015assertions} conducted a study to investigate the correlation between assertions and test effectiveness based on Java programs, and found that both assertion numbers and assertion coverage are strongly correlated with test effectiveness.

Different from them, our study is to investigate the correlation between test coverage and test effectiveness in \textit{DNN testing}.
Traditional software testing and DNN testing have different test coverage and their subjects have totally different characteristics.

\printendnotes

\bibliography{sample}

\begin{thebibliography}{73}
\providecommand{\natexlab}[1]{#1}
\providecommand{\url}[1]{\texttt{#1}}
\providecommand{\urlprefix}{}

\bibitem[{Chen et~al.(2015)Chen, Chenyi and Seff, Ari and Kornhauser, Alain and
  Xiao, Jianxiong}]{chen2015deepdriving}
Chen C, Seff A, Kornhauser A, Xiao J.
\newblock Deepdriving: Learning affordance for direct perception in autonomous
  driving.
\newblock In: ICCV; 2015. p. 2722--2730.

\bibitem[{Sun et~al.(2014)Yi Sun and Yuheng Chen and Xiaogang Wang and Xiaoou
  Tang}]{sun2014deep}
Sun Y, Chen Y, Wang X, Tang X.
\newblock Deep Learning Face Representation by Joint
  Identification-Verification.
\newblock In: {NeurIPS}; 2014. p. 1988--1996.

\bibitem[{Cheng(2019)Cheng, Yong}]{cheng2019semi}
Cheng Y.
\newblock Semi-supervised learning for neural machine translation.
\newblock In: Joint Training for Neural Machine Translation; 2019.p. 25--40.

\bibitem[{Obermeyer and Emanuel(2016)Obermeyer, Ziad and Emanuel, Ezekiel
  J}]{obermeyer2016predicting}
Obermeyer Z, Emanuel EJ.
\newblock Predicting the future—big data, machine learning, and clinical
  medicine.
\newblock N Engl J Med 2016;375(13):1216.

\bibitem[{Gu et~al.(2018)Xiaodong Gu and Hongyu Zhang and Sunghun
  Kim}]{DBLP:conf/icse/GuZ018}
Gu X, Zhang H, Kim S.
\newblock Deep code search.
\newblock In: ICSE; 2018. p. 933--944.

\bibitem[{Gu et~al.(2016)Xiaodong Gu and Hongyu Zhang and Dongmei Zhang and
  Sunghun Kim}]{DBLP:conf/sigsoft/GuZZK16}
Gu X, Zhang H, Zhang D, Kim S.
\newblock Deep {API} learning.
\newblock In: FSE; 2016. p. 631--642.

\bibitem[{Kang et~al.(2021)Yuning Kang and Zan Wang and Hongyu Zhang and Junjie
  Chen and Hanmo You}]{kang2021APIRecX}
Kang Y, Wang Z, Zhang H, Chen J, You H.
\newblock APIRecX: Cross-Library {API} Recommendation via Pre-Trained Language
  Model.
\newblock In: {EMNLP} {(1)} Association for Computational Linguistics; 2021. p.
  3425--3436.

\bibitem[{Tian et~al.(2022)Zhao Tian and Junjie Chen and Qihao Zhu and Junjie
  Yang and Lingming Zhang}]{tian2022learning}
Tian Z, Chen J, Zhu Q, Yang J, Zhang L.
\newblock Learning to Construct Better Mutation Faults.
\newblock In: {ASE} {ACM}; 2022. p. 64:1--64:13.

\bibitem[{Pei et~al.(2017)Pei, Kexin and Cao, Yinzhi and Yang, Junfeng and
  Jana, Suman}]{pei2017deepxplore}
Pei K, Cao Y, Yang J, Jana S.
\newblock Deepxplore: Automated whitebox testing of deep learning systems.
\newblock In: proceedings of the 26th Symposium on Operating Systems
  Principles; 2017. p. 1--18.

\bibitem[{Ma et~al.(2018)Ma, Lei and Juefei-Xu, Felix and Zhang, Fuyuan and
  Sun, Jiyuan and Xue, Minhui and Li, Bo and Chen, Chunyang and Su, Ting and
  Li, Li and Liu, Yang and others}]{ma2018deepgauge}
Ma L, Juefei-Xu F, Zhang F, Sun J, Xue M, Li B, et~al.
\newblock Deepgauge: Multi-granularity testing criteria for deep learning
  systems.
\newblock In: ASE; 2018. p. 120--131.

\bibitem[{Odena et~al.(2019)Odena, Augustus and Olsson, Catherine and Andersen,
  David and Goodfellow, Ian}]{odena2018tensorfuzz}
Odena A, Olsson C, Andersen D, Goodfellow I.
\newblock Tensorfuzz: Debugging neural networks with coverage-guided fuzzing.
\newblock In: ICML; 2019. p. 4901--4911.

\bibitem[{new(Accessed: 2022)}]{news1}
News; Accessed: 2022.
\newblock
  \url{https://www.vice.com/en_us/article/9kga85/uber-is-giving-up-on-self-driving-cars-in-california-after-deadly-crash}.

\bibitem[{Inozemtseva and Holmes(2014)Inozemtseva, Laura and Holmes,
  Reid}]{inozemtseva2014coverage}
Inozemtseva L, Holmes R.
\newblock Coverage is not strongly correlated with test suite effectiveness.
\newblock In: ICSE; 2014. p. 435--445.

\bibitem[{Zhang and Mesbah(2015)Zhang, Yucheng and Mesbah,
  Ali}]{zhang2015assertions}
Zhang Y, Mesbah A.
\newblock Assertions are strongly correlated with test suite effectiveness.
\newblock In: FSE; 2015. p. 214--224.

\bibitem[{Chekam et~al.(2017)Thierry Titcheu Chekam and Mike Papadakis and Yves
  Le Traon and Mark Harman}]{DBLP:conf/icse/ChekamPTH17}
Chekam TT, Papadakis M, Traon YL, Harman M.
\newblock An empirical study on mutation, statement and branch coverage fault
  revelation that avoids the unreliable clean program assumption.
\newblock In: ICSE; 2017. p. 597--608.

\bibitem[{Morrison et~al.(2012)G. C. Morrison and Cornelia P. Inggs and W. C.
  Visser}]{DBLP:conf/saicsit/MorrisonIV12}
Morrison GC, Inggs CP, Visser WC.
\newblock Automated coverage calculation and test case generation.
\newblock In: {SAICSIT} {ACM}; 2012. p. 84--93.

\bibitem[{Hilton et~al.(2018)Michael Hilton and Jonathan Bell and Darko
  Marinov}]{DBLP:conf/kbse/Hilton0M18}
Hilton M, Bell J, Marinov D.
\newblock A large-scale study of test coverage evolution.
\newblock In: ASE; 2018. p. 53--63.

\bibitem[{Gligoric et~al.(2013)Milos Gligoric and Alex Groce and Chaoqiang
  Zhang and Rohan Sharma and Mohammad Amin Alipour and Darko
  Marinov}]{DBLP:conf/issta/GligoricGZSAM13}
Gligoric M, Groce A, Zhang C, Sharma R, Alipour MA, Marinov D.
\newblock Comparing non-adequate test suites using coverage criteria.
\newblock In: ISSTA; 2013. p. 302--313.

\bibitem[{Gay et~al.(2016)Gregory Gay and Ajitha Rajan and Matt Staats and
  Michael W. Whalen and Mats Per Erik Heimdahl}]{DBLP:journals/tosem/GayRSWH16}
Gay G, Rajan A, Staats M, Whalen MW, Heimdahl MPE.
\newblock The Effect of Program and Model Structure on the Effectiveness of
  {MC/DC} Test Adequacy Coverage.
\newblock TOSEM 2016;25(3):25:1--25:34.

\bibitem[{Gligoric et~al.(2015)Milos Gligoric and Alex Groce and Chaoqiang
  Zhang and Rohan Sharma and Mohammad Amin Alipour and Darko
  Marinov}]{DBLP:journals/tosem/GligoricGZSAM15}
Gligoric M, Groce A, Zhang C, Sharma R, Alipour MA, Marinov D.
\newblock Guidelines for Coverage-Based Comparisons of Non-Adequate Test
  Suites.
\newblock TOSEM 2015;24(4):22:1--22:33.

\bibitem[{Kim et~al.(2019)Kim, Jinhan and Feldt, Robert and Yoo,
  Shin}]{kim2019guiding}
Kim J, Feldt R, Yoo S.
\newblock Guiding deep learning system testing using surprise adequacy.
\newblock In: ICSE; 2019. p. 1039--1049.

\bibitem[{Carlini and Wagner(2017)Nicholas Carlini and David A.
  Wagner}]{DBLP:conf/sp/Carlini017}
Carlini N, Wagner DA.
\newblock Towards Evaluating the Robustness of Neural Networks.
\newblock In: S\&P; 2017. p. 39--57.

\bibitem[{Harel-Canada et~al.(2020)Harel-Canada, Fabrice and Wang, Lingxiao and
  Gulzar, Muhammad Ali and Gu, Quanquan and Kim, Miryung}]{harel2020neuron}
Harel-Canada F, Wang L, Gulzar MA, Gu Q, Kim M.
\newblock Is neuron coverage a meaningful measure for testing deep neural
  networks?
\newblock In: Proceedings of the 28th ACM Joint Meeting on European Software
  Engineering Conference and Symposium on the Foundations of Software
  Engineering; 2020. p. 851--862.

\bibitem[{Yan et~al.(2020)Yan, Shenao and Tao, Guanhong and Liu, Xuwei and
  Zhai, Juan and Ma, Shiqing and Xu, Lei and Zhang,
  Xiangyu}]{yan2020correlations}
Yan S, Tao G, Liu X, Zhai J, Ma S, Xu L, et~al.
\newblock Correlations between deep neural network model coverage criteria and
  model quality.
\newblock In: Proceedings of the 28th ACM Joint Meeting on European Software
  Engineering Conference and Symposium on the Foundations of Software
  Engineering; 2020. p. 775--787.

\bibitem[{Yang et~al.(2022)Yang, Zhou and Shi, Jieke and Asyrofi, Muhammad
  Hilmi and Lo, David}]{yang2022revisiting}
Yang Z, Shi J, Asyrofi MH, Lo D.
\newblock Revisiting Neuron Coverage Metrics and Quality of Deep Neural
  Networks.
\newblock arXiv preprint arXiv:220100191 2022;.

\bibitem[{Gerasimou et~al.(2020)Gerasimou, Simos and Eniser, Hasan Ferit and
  Sen, Alper and Cakan, Alper}]{gerasimou2020importance}
Gerasimou S, Eniser HF, Sen A, Cakan A.
\newblock Importance-driven deep learning system testing.
\newblock In: ICSE IEEE; 2020. p. 702--713.

\bibitem[{hom(Accessed: 2022)}]{homepage}
Homepage; Accessed: 2022.
\newblock \url{https://github.com/Jacob-yen/DL-Coverage-Study}.

\bibitem[{Bach et~al.(2015)Bach, Sebastian and Binder, Alexander and Montavon,
  Gr{\'e}goire and Klauschen, Frederick and M{\"u}ller, Klaus-Robert and Samek,
  Wojciech}]{bach2015pixel}
Bach S, Binder A, Montavon G, Klauschen F, M{\"u}ller KR, Samek W.
\newblock On pixel-wise explanations for non-linear classifier decisions by
  layer-wise relevance propagation.
\newblock PloS one 2015;10(7):e0130140.

\bibitem[{Wand and Jones(1994)Wand, Matt P and Jones, M Chris}]{wand1994kernel}
Wand MP, Jones MC.
\newblock Kernel smoothing.
\newblock Chapman and Hall/CRC; 1994.

\bibitem[{mni(Accessed: 2022)}]{mnist}
MNIST; Accessed: 2022.
\newblock \url{http://yann.lecun.com/exdb/mnist/}.

\bibitem[{Xiao et~al.(2017)Han Xiao and Kashif Rasul and Roland
  Vollgraf}]{xiao2017/online}
Xiao H, Rasul K, Vollgraf R, Fashion-MNIST: a Novel Image Dataset for
  Benchmarking Machine Learning Algorithms; 2017.

\bibitem[{cif(Accessed: 2022)}]{cifar}
CIFAR-10; Accessed: 2022.
\newblock \url{http://www.cs.toronto.edu/~kriz/cifar.html}.

\bibitem[{dri(Accessed: 2022)}]{driving}
Driving; Accessed: 2022.
\newblock \url{https://udacity.com/self-driving-car}.

\bibitem[{spe(Accessed: 2022)}]{speech}
Speech-Commands; Accessed: 2022.
\newblock
  \url{https://github.com/bjtommychen/Keras_DeepSpeech2_SpeechRecognition}.

\bibitem[{20n(Accessed: 2022)}]{20news}
20-Newsgroups; Accessed: 2022.
\newblock \url{http://qwone.com/~jason/20Newsgroups/}.

\bibitem[{Li et~al.(2019)Li, Zenan and Ma, Xiaoxing and Xu, Chang and Cao, Chun
  and Xu, Jingwei and L\"{u}, Jian}]{Li:2019:BOD:3338906.3338930}
Li Z, Ma X, Xu C, Cao C, Xu J, L\"{u} J.
\newblock Boosting Operational DNN Testing Efficiency Through Conditioning.
\newblock In: FSE; 2019. p. 499--509.

\bibitem[{Chen et~al.(2020)Chen, Junjie and Wu, Zhuo and Wang, Zan and You,
  Hanmo and Zhang, Lingming and Yan, Ming}]{chenpracticalTOSEM}
Chen J, Wu Z, Wang Z, You H, Zhang L, Yan M.
\newblock Practical Accuracy Estimation for Efficient Deep Neural Network
  Testing.
\newblock TOSEM 2020;29(4):30:1--30:35.

\bibitem[{Myers and Sirois(2004)Myers, Leann and Sirois, Maria
  J}]{myers2004spearman}
Myers L, Sirois MJ.
\newblock Spearman correlation coefficients, differences between.
\newblock Encyclopedia of statistical sciences 2004;12.

\bibitem[{Chen et~al.(2017)Chen, Junjie and Bai, Yanwei and Hao, Dan and Zhang,
  Lingming and Zhang, Lu and Xie, Bing}]{chen2017assertions}
Chen J, Bai Y, Hao D, Zhang L, Zhang L, Xie B.
\newblock How do assertions impact coverage-based test-suite reduction?
\newblock In: ICST; 2017. p. 418--423.

\bibitem[{Goodfellow et~al.(2015)Ian J. Goodfellow and Jonathon Shlens and
  Christian Szegedy}]{DBLP:journals/corr/GoodfellowSS14}
Goodfellow IJ, Shlens J, Szegedy C.
\newblock Explaining and Harnessing Adversarial Examples.
\newblock In: ICLR; 2015. .

\bibitem[{Kurakin et~al.(2017)Alexey Kurakin and Ian J. Goodfellow and Samy
  Bengio}]{DBLP:conf/iclr/KurakinGB17a}
Kurakin A, Goodfellow IJ, Bengio S.
\newblock Adversarial examples in the physical world.
\newblock In: ICLR; 2017. .

\bibitem[{Papernot et~al.(2016)Nicolas Papernot and Patrick D. McDaniel and
  Somesh Jha and Matt Fredrikson and Z. Berkay Celik and Ananthram
  Swami}]{DBLP:conf/eurosp/PapernotMJFCS16}
Papernot N, McDaniel PD, Jha S, Fredrikson M, Celik ZB, Swami A.
\newblock The Limitations of Deep Learning in Adversarial Settings.
\newblock In: S\&P; 2016. p. 372--387.

\bibitem[{Carlini and Wagner(2018)Nicholas Carlini and David A.
  Wagner}]{DBLP:conf/sp/Carlini018}
Carlini N, Wagner DA.
\newblock Audio Adversarial Examples: Targeted Attacks on Speech-to-Text.
\newblock In: S\&P Workshops; 2018. p. 1--7.

\bibitem[{Ren et~al.(2019)Shuhuai Ren and Yihe Deng and Kun He and Wanxiang
  Che}]{pwws2019}
Ren S, Deng Y, He K, Che W.
\newblock Generating Natural Language Adversarial Examples through Probability
  Weighted Word Saliency.
\newblock In: {ACL} {(1)} Association for Computational Linguistics; 2019. p.
  1085--1097.

\bibitem[{Sun et~al.(2019{\natexlab{a}})Youcheng Sun and Xiaowei Huang and
  Daniel Kroening and James Sharp and Matthew Hill and Rob
  Ashmore}]{DBLP:journals/tecs/SunHKSHA19}
Sun Y, Huang X, Kroening D, Sharp J, Hill M, Ashmore R.
\newblock Structural Test Coverage Criteria for Deep Neural Networks.
\newblock TECS 2019;18(5s):94:1--94:23.

\bibitem[{Sun et~al.(2019{\natexlab{b}})Youcheng Sun and Xiaowei Huang and
  Daniel Kroening and James Sharp and Matthew Hill and Rob
  Ashmore}]{DBLP:conf/icse/SunHKSHA19}
Sun Y, Huang X, Kroening D, Sharp J, Hill M, Ashmore R.
\newblock DeepConcolic: testing and debugging deep neural networks.
\newblock In: ICSE; 2019. p. 111--114.

\bibitem[{Zhou et~al.(2021)Zhou, Zhiyang and Dou, Wensheng and Liu, Jie and
  Zhang, Chenxin and Wei, Jun and Ye, Dan}]{zhou2021deepcon}
Zhou Z, Dou W, Liu J, Zhang C, Wei J, Ye D.
\newblock DeepCon: Contribution Coverage Testing for Deep Learning Systems.
\newblock In: 2021 IEEE International Conference on Software Analysis,
  Evolution and Reengineering (SANER) IEEE; 2021. p. 189--200.

\bibitem[{Ma et~al.(2019)Lei Ma and Felix Juefei{-}Xu and Minhui Xue and Bo Li
  and Li Li and Yang Liu and Jianjun Zhao}]{ma2019deepct}
Ma L, Juefei{-}Xu F, Xue M, Li B, Li L, Liu Y, et~al.
\newblock DeepCT: Tomographic Combinatorial Testing for Deep Learning Systems.
\newblock In: {SANER}; 2019. p. 614--618.

\bibitem[{Du et~al.(2019)Xiaoning Du and Xiaofei Xie and Yi Li and Lei Ma and
  Yang Liu and Jianjun Zhao}]{du2018deepcruiser}
Du X, Xie X, Li Y, Ma L, Liu Y, Zhao J.
\newblock DeepStellar: model-based quantitative analysis of stateful deep
  learning systems.
\newblock In: {ESEC/SIGSOFT} {FSE}; 2019. p. 477--487.

\bibitem[{Ma et~al.(2018)Shiqing Ma and Yingqi Liu and Wen{-}Chuan Lee and
  Xiangyu Zhang and Ananth Grama}]{DBLP:conf/sigsoft/MaLLZG18}
Ma S, Liu Y, Lee W, Zhang X, Grama A.
\newblock {MODE:} automated neural network model debugging via state
  differential analysis and input selection.
\newblock In: FSE; 2018. p. 175--186.

\bibitem[{Sun et~al.(2018)Sun, Youcheng and Huang, Xiaowei and Kroening,
  Daniel}]{sun2018testing}
Sun Y, Huang X, Kroening D.
\newblock Testing deep neural networks.
\newblock arXiv preprint arXiv:180304792 2018;.

\bibitem[{Zhang et~al.(2022)Zhang, Jie M and Harman, Mark and Ma, Lei and Liu,
  Yang}]{zhang2019machine}
Zhang JM, Harman M, Ma L, Liu Y.
\newblock Machine learning testing: Survey, landscapes and horizons.
\newblock TSE 2022;48(2):1--36.

\bibitem[{Zhang et~al.(2019)Fuyuan Zhang and Sankalan Pal Chowdhury and Maria
  Christakis}]{DBLP:journals/corr/abs-1910-06296}
Zhang F, Chowdhury SP, Christakis M.
\newblock DeepSearch: Simple and Effective Blackbox Fuzzing of Deep Neural
  Networks.
\newblock CoRR 2019;abs/1910.06296.

\bibitem[{Tian et~al.(2018)Yuchi Tian and Kexin Pei and Suman Jana and
  Baishakhi Ray}]{DBLP:conf/icse/TianPJR18}
Tian Y, Pei K, Jana S, Ray B.
\newblock DeepTest: automated testing of deep-neural-network-driven autonomous
  cars.
\newblock In: ICSE; 2018. p. 303--314.

\bibitem[{Lee et~al.(2020)Seokhyun Lee and Sooyoung Cha and Dain Lee and Hakjoo
  Oh}]{DBLP:conf/issta/LeeCLO20}
Lee S, Cha S, Lee D, Oh H.
\newblock Effective white-box testing of deep neural networks with adaptive
  neuron-selection strategy.
\newblock In: ISSTA; 2020. p. 165--176.

\bibitem[{Ma et~al.(2018)Ma, Lei and Zhang, Fuyuan and Sun, Jiyuan and Xue,
  Minhui and Li, Bo and Juefei-Xu, Felix and Xie, Chao and Li, Li and Liu, Yang
  and Zhao, Jianjun and others}]{ma2018deepmutation}
Ma L, Zhang F, Sun J, Xue M, Li B, Juefei-Xu F, et~al.
\newblock Deepmutation: Mutation testing of deep learning systems.
\newblock In: ISSRE; 2018. p. 100--111.

\bibitem[{Li et~al.(2019)Zenan Li and Xiaoxing Ma and Chang Xu and Chun
  Cao}]{li2019structural}
Li Z, Ma X, Xu C, Cao C.
\newblock Structural coverage criteria for neural networks could be misleading.
\newblock In: {ICSE} {(NIER)}; 2019. p. 89--92.

\bibitem[{Dong et~al.(2020)Dong, Yizhen and Zhang, Peixin and Wang, Jingyi and
  Liu, Shuang and Sun, Jun and Hao, Jianye and Wang, Xinyu and Wang, Li and
  Dong, Jinsong and Dai, Ting}]{dong2020empirical}
Dong Y, Zhang P, Wang J, Liu S, Sun J, Hao J, et~al.
\newblock An Empirical Study on Correlation between Coverage and Robustness for
  Deep Neural Networks.
\newblock In: 2020 25th International Conference on Engineering of Complex
  Computer Systems (ICECCS) IEEE; 2020. p. 73--82.

\bibitem[{Bastani et~al.(2016)Bastani, Osbert and Ioannou, Yani and
  Lampropoulos, Leonidas and Vytiniotis, Dimitrios and Nori, Aditya and
  Criminisi, Antonio}]{bastani2016measuring}
Bastani O, Ioannou Y, Lampropoulos L, Vytiniotis D, Nori A, Criminisi A.
\newblock Measuring neural net robustness with constraints.
\newblock In: NeurIPS; 2016. p. 2613--2621.

\bibitem[{Jha et~al.(2019)Susmit Jha and Sunny Raj and Steven Lawrence
  Fernandes and Sumit Kumar Jha and Somesh Jha and Brian Jalaian and Gunjan
  Verma and Ananthram Swami}]{jha2019attribution}
Jha S, Raj S, Fernandes SL, Jha SK, Jha S, Jalaian B, et~al.
\newblock Attribution-Based Confidence Metric For Deep Neural Networks.
\newblock In: NeurIPS; 2019. p. 11826--11837.

\bibitem[{Gopinath et~al.(2018)Divya Gopinath and Guy Katz and Corina S.
  Pasareanu and Clark W. Barrett}]{gopinath2018deepsafe}
Gopinath D, Katz G, Pasareanu CS, Barrett CW.
\newblock DeepSafe: {A} Data-Driven Approach for Assessing Robustness of Neural
  Networks.
\newblock In: {ATVA}, vol. 11138; 2018. p. 3--19.

\bibitem[{Katz et~al.(2017)Guy Katz and Clark W. Barrett and David L. Dill and
  Kyle Julian and Mykel J. Kochenderfer}]{katz2017towards}
Katz G, Barrett CW, Dill DL, Julian K, Kochenderfer MJ.
\newblock Towards Proving the Adversarial Robustness of Deep Neural Networks.
\newblock In: FVAV@iFM, vol. 257 of {EPTCS}; 2017. p. 19--26.

\bibitem[{Guo et~al.(2018)Guo, Jianmin and Jiang, Yu and Zhao, Yue and Chen,
  Quan and Sun, Jiaguang}]{guo2018dlfuzz}
Guo J, Jiang Y, Zhao Y, Chen Q, Sun J.
\newblock Dlfuzz: Differential fuzzing testing of deep learning systems.
\newblock In: FSE; 2018. p. 739--743.

\bibitem[{Xie et~al.(2019)Xie, Xiaofei and Ma, Lei and Juefei-Xu, Felix and
  Xue, Minhui and Chen, Hongxu and Liu, Yang and Zhao, Jianjun and Li, Bo and
  Yin, Jianxiong and See, Simon}]{xie2019deephunter}
Xie X, Ma L, Juefei-Xu F, Xue M, Chen H, Liu Y, et~al.
\newblock DeepHunter: a coverage-guided fuzz testing framework for deep neural
  networks.
\newblock In: ISSTA; 2019. p. 146--157.

\bibitem[{Sun et~al.(2018)Sun, Youcheng and Wu, Min and Ruan, Wenjie and Huang,
  Xiaowei and Kwiatkowska, Marta and Kroening, Daniel}]{sun2018concolic}
Sun Y, Wu M, Ruan W, Huang X, Kwiatkowska M, Kroening D.
\newblock Concolic testing for deep neural networks.
\newblock In: ASE; 2018. p. 109--119.

\bibitem[{Shen et~al.(2022)Qingchao Shen and Junjie Chen and Jie M. Zhang and
  Haoyu Wang and Shuang Liu and Menghan Tian}]{shen2022QAQA}
Shen Q, Chen J, Zhang JM, Wang H, Liu S, Tian M.
\newblock Natural Test Generation for Precise Testing of Question Answering
  Software.
\newblock In: {ASE} {ACM}; 2022. p. 71:1--71:12.

\bibitem[{You et~al.(2023)You, Hanmo and Wang, Zan and Chen, Junjie and Liu,
  Shuang and Li, Shuochuan}]{you2023DRFuzz}
You H, Wang Z, Chen J, Liu S, Li S.
\newblock Regression Fuzzing for Deep Learning Systems.
\newblock In: 45th International Conference on Software Engineering; 2023. To
  appear.

\bibitem[{Wang et~al.(2020)Zan Wang and Ming Yan and Junjie Chen and Shuang Liu
  and Dongdi Zhang}]{wang2020lemon}
Wang Z, Yan M, Chen J, Liu S, Zhang D.
\newblock Deep learning library testing via effective model generation.
\newblock In: {ESEC/SIGSOFT} {FSE} {ACM}; 2020. p. 788--799.

\bibitem[{Zhang et~al.(2020)Yuhao Zhang and Luyao Ren and Liqian Chen and
  Yingfei Xiong and Shing{-}Chi Cheung and Tao Xie}]{zhang2020DEBAR}
Zhang Y, Ren L, Chen L, Xiong Y, Cheung S, Xie T.
\newblock Detecting numerical bugs in neural network architectures.
\newblock In: {ESEC/SIGSOFT} {FSE} {ACM}; 2020. p. 826--837.

\bibitem[{Yan et~al.(2021)Ming Yan and Junjie Chen and Xiangyu Zhang and Lin
  Tan and Gan Wang and Zan Wang}]{yan2021grist}
Yan M, Chen J, Zhang X, Tan L, Wang G, Wang Z.
\newblock Exposing numerical bugs in deep learning via gradient
  back-propagation.
\newblock In: {ESEC/SIGSOFT} {FSE} {ACM}; 2021. p. 627--638.

\bibitem[{Ma et~al.(2019)Wei Ma and Mike Papadakis and Anestis Tsakmalis and
  Maxime Cordy and Yves Le Traon}]{DBLP:journals/corr/abs-1904-13195}
Ma W, Papadakis M, Tsakmalis A, Cordy M, Traon YL.
\newblock Test Selection for Deep Learning Systems.
\newblock CoRR 2019;abs/1904.13195.

\bibitem[{Feng et~al.(2020)Yang Feng and Qingkai Shi and Xinyu Gao and Jun Wan
  and Chunrong Fang and Zhenyu Chen}]{shi2019deepgini}
Feng Y, Shi Q, Gao X, Wan J, Fang C, Chen Z.
\newblock DeepGini: prioritizing massive tests to enhance the robustness of
  deep neural networks.
\newblock In: {ISSTA}; 2020. p. 177--188.

\bibitem[{Namin and Andrews(2009)Namin, Akbar Siami and Andrews, James
  H}]{namin2009influence}
Namin AS, Andrews JH.
\newblock The influence of size and coverage on test suite effectiveness.
\newblock In: ISSTA; 2009. p. 57--68.

\end{thebibliography}



\end{document}